\documentclass[hidelinks,onefignum,onetabnum]{siamart251216}
\usepackage{hyperref}
\usepackage{amsfonts,amssymb,amsmath}
\usepackage{enumitem}
\usepackage{pifont}
\usepackage{booktabs}       
\usepackage{multirow}
\usepackage{mathtools}
\usepackage{makecell}
\usepackage{algorithmic}
\usepackage{placeins}
\usepackage{xcolor}
\def\etal{\textit{et al.}}

\title{Reconstructing Interacting Blood Flow and Wall Interface Deformation from Noisy Flow Data Using PirateNets
}

\author{
Han Zhang\thanks{Division of Applied Mathematics, Brown University, RI 02906, USA.
(\email{han\_zhang6@brown.edu})}
\and Xue-Cheng Tai\thanks{Norwegian Research Centre (NORCE), Nygardsgaten 112, 5008 Bergen, Norway. 
(\email{xtai@norceresearch.no})}
\and Alessandro Veneziani\thanks{The author is deceased. Former address: Department of Mathematics and Computer Science, Emory University, GA 30322, USA.}
\and Raymond H. Chan\thanks{Department of Operations and Risk Management and the School of Data Science, Lingnan University and Hong Kong Centre for Cerebro-Cardiovascular Health Engineering, Hong Kong, China. 
(\email{raymond.chan@ln.edu.hk})}
}

\begin{document}

\maketitle
\begin{abstract}
Blood flow and elastic vessel-wall motion interact with each other: fluid stress drives wall deformation, while the moving wall changes the domain in which blood flows. Given noisy flow images, we study the inverse problem of reconstructing the coupled, unsteady three-dimensional motions. We solve the reconstruction through Gauss--Seidel-type alternating fluid and geometry updates. For the fluid, we use physics-informed networks to recover velocity and pressure by fitting the observations while enforcing the incompressible Navier--Stokes equations on the moving domain, which is described by an arbitrary Lagrangian--Eulerian (ALE) map. For the geometry, a reference-domain correction and sequential registration recover the vessel shape and wall displacement. Interface traction continuity and an elastic momentum equation regularize the registration so that the recovered displacement is also physics-informed. The networks evaluate fields and physical residuals at sampled space--time points without remeshing the moving domain. Experiments on a converging channel and a patient-like aorta demonstrate reconstruction of flow and wall motion from noisy data. Ablation and parameter-sensitivity studies examine the role of structure regularization and imperfect physical coefficients. A further experiment with a three-element Windkessel outlet condition demonstrates how the framework accommodates a more physiologically realistic boundary setting. 
\end{abstract}



\section{Introduction}
\label{sec:introduction}

Quantitative blood-flow measurements provide information about cardiovascular function that supports disease monitoring, diagnosis, and treatment planning. Velocity imaging is an important way to obtain such data. In particular, phase-contrast magnetic resonance imaging, including four-dimensional flow MRI, measures time-dependent velocity in three-dimensional vascular regions~\cite{markl20124d}.

However, acquiring high-quality flow images remains challenging. Long acquisition times increase sensitivity to patient motion~\cite{gigengack2011motion}, whereas faster acquisitions may reduce spatial resolution and signal quality. Device-related errors can distort measured velocities, and noise and limited resolution can make vessel boundaries difficult to identify~\cite{markl20124d,ruthotto2012diffeomorphic}. As a result, uncertainty in both the observed flow and the vessel geometry may make the data unsuitable for direct use.

Physics-based inverse methods provide a way to recover information that velocity measurements alone cannot determine. Aguayo~\etal~\cite{aguayo2021distributed,aguayo2022analysis,aguayo2022stability,aguayo2023distributed} and Kontogiannis~\etal~\cite{kontogiannis2022joint,kontogiannis2024bayesian} infer flow obstacles, boundaries, or model parameters from velocity data. Zhang~\etal~\cite{zhang2026coupled} jointly reconstruct steady two-dimensional flow and lumen geometry. Time-dependent inverse methods recover flow without jointly reconstructing elastic wall motion and geometry~\cite{raissi2020hidden}, whereas inverse fluid--structure interaction methods model wall deformation but focus on parameter estimation from a prescribed reference geometry~\cite{bertagna2014data,bertoglio2012sequential,perego2011variational}. To our knowledge, reconstructing coupled unsteady three-dimensional flow and elastic vessel-wall motion together from noisy velocity images remains unexplored; see~\cite{nolte2022inverse} for a review.

We consider the reconstruction of unsteady three-dimensional blood flow and vessel-wall motion from noisy, time-dependent velocity images. We start from an approximate reference-domain mask guess with the same topology as the target vessel~\cite{zhang2026coupled}. Fluid viscosity, wall density, and elastic parameters are supplied, but may be imperfect in some tests. The unknowns are the corrected reference geometry, the evolving lumen and wall displacement, and the fluid velocity and pressure. Neither the inlet trace $\boldsymbol{g}$ nor the initial velocity $\boldsymbol{u}_0$ is prescribed; both are represented implicitly by the reconstructed velocity field.

We describe the moving domain with an arbitrary Lagrangian--Eulerian (ALE) map and accomplish this task through Gauss--Seidel-type alternating between fluid and geometry updates. The fluid update enforces the incompressible Navier--Stokes equations. The geometry update combines diffeomorphic registration~\cite{mang2017lagrangian,burger2013hyperelastic} with fluid--structure traction continuity and an interface-evaluated elastic momentum residual, which help reduce ambiguity among material-point motions producing similar deformed lumens. The physics-informed network architecture PirateNets~\cite{wang2024piratenets} is used to evaluate the fields and residuals at sampled space--time points without remeshing the moving domain.

The contributions are as follows:
\begin{itemize}[leftmargin=*]
  \item We formulate an inverse problem for coupled, unsteady three-dimensional blood flow and vessel-wall motion, with incompressible Navier--Stokes flow in a moving domain and wall kinematics linking the fluid to the vessel.
  \item We formulate the geometry update as sequential hyperelastic registration, using traction continuity and an interface-evaluated solid momentum residual to regularize wall displacement. The registration maps provide image-driven domain updates analogous to shape-gradient steps.
  \item We develop a Gauss--Seidel-type alternating algorithm using PirateNets, whose ability to represent high-frequency features helps reconstruct spatially varying flow from noisy velocity images.
  \item We extend the framework to a three-element Windkessel outlet model, providing a more physiologically realistic boundary condition and bringing the reconstruction setting closer to clinical blood flow.
\end{itemize}


\section{Related Work}
\label{sec:related}

\subsection{Forward Simulation of Blood Flow}
Forward blood-flow models compute velocity and pressure from prescribed geometry, physical parameters, and initial and boundary data. Quarteroni~\etal~\cite{quarteroni2000computational,quarteroni2004mathematical} describe cardiovascular flow modeling, while Formaggia~\etal~\cite{formaggia2001coupling,formaggia2010flowrate} address multiscale coupling and flow-rate conditions in compliant vessels. For deformable walls, Figueroa~\etal~\cite{figueroa2006coupled} couple wall momentum to three-dimensional blood flow. Monolithic and ALE domain-decomposition solvers are developed by Barker and Cai~\cite{barker2010scalable} and Wu and Cai~\cite{wu2014fully}; parallel and interface-operator algorithms are studied by Crosetto~\etal~\cite{crosetto2011parallel} and Deparis~\etal~\cite{deparis2006steklov}. Liu~\etal~\cite{liu2023two,liu2024coarse} develop multilevel preconditioners for three-dimensional arterial Stokes and Navier--Stokes flows.

Partitioned coupling and added-mass effects are treated by Causin~\etal~\cite{causin2005addedmass}, Quaini~\etal~\cite{quaini2007semi,quaini2012validation}, and Badia~\etal~\cite{badia2008splitting,badia2008modular,badia2009coupling,badia2008robin}. Nobile and Vergara~\cite{nobile2008effective} derive a vascular FSI formulation with generalized Robin conditions. Other treatments of moving vessels and elastic structures include those of Buka\v{c}~\etal~\cite{bukavc2013fluid}, and Basting~\etal~\cite{basting2017extended}. Neural forward solvers use physics-informed residuals for incompressible flow~\cite{raissi2019physics}. Zhang~\etal~\cite{zhang2024meshless,zhang2024full,zhang2025parametrized} extend meshless neural simulation to elastic vessels, including fully three-dimensional curved geometries.

\subsection{Inverse Reconstruction of Flow Images}
Inverse flow methods recover unknown fields or domains from measured velocity. Aguayo~\etal~\cite{aguayo2021distributed,aguayo2022analysis,aguayo2022stability,aguayo2023distributed} study obstacle and permeability identification using distributed-resistance and related Navier--Stokes formulations. Kontogiannis~\etal~\cite{kontogiannis2022joint,kontogiannis2024bayesian} combine velocity reconstruction with boundary or parameter inference from noisy flow data. Vascular shape optimization was also studied by Quarteroni and Rozza~\cite{quarteroni2003optimal}. Zhang~\etal~\cite{zhang2026coupled} reconstruct steady two-dimensional flow and lumen geometry from noisy images by alternating a physics-informed fluid solve with quasi-conformal registration. 

Time-dependent variational and physics-informed methods reconstruct pulsatile flow from velocity measurements or flow visualizations, but do not jointly infer elastic vessel-wall motion and the evolving geometry~\cite{raissi2020hidden}. Inverse fluid--structure interaction methods account for wall deformation, but primarily estimate material or boundary parameters from measurements on a prescribed reference geometry~\cite{bertagna2014data,bertoglio2012sequential,perego2011variational}. See~\cite{nolte2022inverse} for a review of inverse blood-flow problems. Fung and Ruthotto~\cite{fung2019multiscale} used multiscale finite-volume model reduction to accelerate PDE-constrained parameter estimation by updating the reduced space during optimization. To our knowledge, the joint reconstruction of unsteady three-dimensional flow and elastic vessel-wall motion from noisy velocity images remains unexplored.

\section{The Reconstruction Problem}
\label{sec:formulation}

The objective of this work is to reconstruct the coupled fluid dynamics and vessel wall interface motion from noisy flow measurements. Given velocity observations obtained from imaging techniques, such as flow MRI, we aim to recover the unknown fluid dynamics, inlet and initial conditions, and the wall interface deformation that generate a physically consistent blood flow.

Table~\ref{tab:problem-notation} summarizes the main notation used in the formulation. Throughout, $\mathrm j\in\{\mathrm i,\mathrm o,\mathrm w\}$ denotes the inlet, outlet, or vessel wall interface.

\begin{table}[htbp]
\centering
\caption{Principal notation for the formulated problem.}
\label{tab:problem-notation}
\begingroup
\small
\setlength{\tabcolsep}{3pt}
\renewcommand{\arraystretch}{1.05}
\begin{tabular}{@{}>{\raggedright\arraybackslash}p{0.16\linewidth}>{\raggedright\arraybackslash}p{0.30\linewidth}@{\hspace{8pt}}>{\raggedright\arraybackslash}p{0.16\linewidth}>{\raggedright\arraybackslash}p{0.30\linewidth}@{}}
\toprule
Symbol & Meaning & Symbol & Meaning \\
\midrule
$D$ & Image domain & $\tilde{\boldsymbol{u}}$ & Observed velocity \\
$\Omega_{\mathrm{ref}},\Omega_0$ & Prior / reference lumen & $\boldsymbol{u},p$ & Velocity / scaled pressure \\
$\Omega_t,\Omega_T$ & Current / space--time & $\boldsymbol{u}_0,\boldsymbol{g}$ & Initial / inlet velocities \\
$\Gamma_0^{\mathrm j},\Gamma_t^{\mathrm j},\Gamma_T^{\mathrm j}$ & Reference / current / space--time boundary& $\nu,\rho_{\mathrm f},\rho_{\mathrm s}$ & Viscosity / fluid and wall densities \\
$\boldsymbol{x}_0,\boldsymbol{x}_t$ & Reference / current points & $\lambda_{\mathrm s},\mu_{\mathrm s}$ & Scaled Lam\'e moduli \\
$\mathcal A_t,\boldsymbol F_t,J_t$ & ALE map / deformation gradient / determinant & $\boldsymbol{\sigma}_{\mathrm f},\boldsymbol{\sigma}_{\mathrm s}$ & Fluid / solid stresses \\
$\boldsymbol{\zeta}$ & Reference correction & $\boldsymbol{\epsilon}_{\mathrm f},\boldsymbol{\epsilon}_{\mathrm s}$ & Fluid / solid strains \\
$\boldsymbol{d}$ & Wall / domain motion & $\boldsymbol{n}_{\mathrm o},\boldsymbol{n}_{\mathrm w}$ & Outlet / wall normals \\
$\boldsymbol{w}$ & Domain velocity & $\mathcal L_{\mathrm{data}},\mathcal L_{\mathrm{str}}$ & Data / structure losses \\
\bottomrule
\end{tabular}
\endgroup
\end{table}

\begin{figure}
    \centering
    \includegraphics[width=0.99\textwidth]{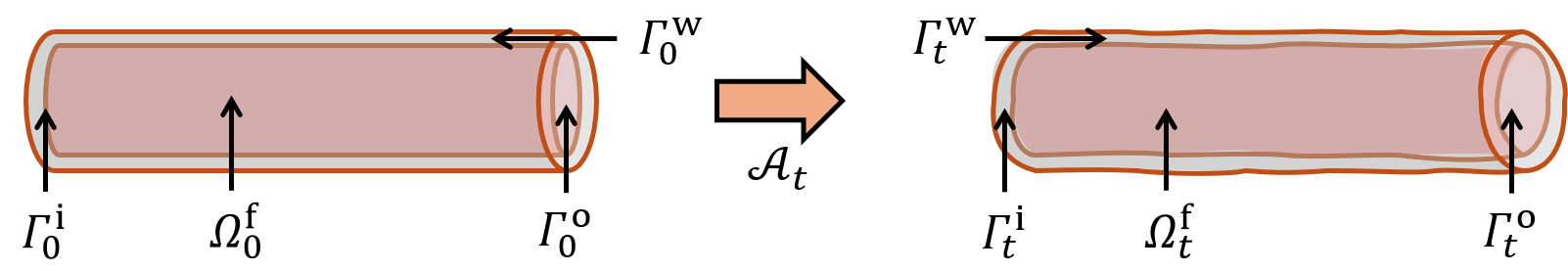}
    \caption{The ALE mapping $\mathcal{A}_t=\boldsymbol{I}+\boldsymbol{d}(\cdot,t)$ carries the
    fixed reference configuration $\Omega_0$, with inlet, outlet, and wall
    boundaries $\Gamma_0^{\mathrm i},\Gamma_0^{\mathrm o},\Gamma_0^{\mathrm w}$,
    onto the time-varying current configuration $\Omega_t$ through the
    displacement field $\boldsymbol{d}(\boldsymbol{x}_0,t)$.}
    \label{fig:alemapping}
\end{figure}

We model blood flow dynamics on a moving three-dimensional domain using an Arbitrary Lagrangian--Eulerian (ALE) mapping~\cite{quarteroni2004mathematical}. In the inverse problem, registration estimates wall displacement, while interface traction continuity and wall-evaluated linear-elastic momentum equation regularize the displacement. Let $D \subset \mathbb{R}^3$ denote the full domain describing the image. We begin with the reference configuration $\Omega_0$, as shown in Figure~\ref{fig:alemapping}. This configuration serves as the parameterization for the entire computational domain, which varies over time. The boundaries of $\Omega_0$ could be divided into the inlet $\Gamma_0^{\mathrm i}$, outlet $\Gamma_0^{\mathrm o}$, and wall interface $\Gamma_0^{\mathrm w}$. Denote each point within the reference configuration domain by the coordinates $\boldsymbol{x}_0=(x_0,y_0,z_0)^\top$. As the domain evolves over time, $\Omega_t$ represents the current configuration at time $t$ with boundaries $\Gamma_t^{\mathrm i}$, $\Gamma_t^{\mathrm o}$, and $\Gamma_t^{\mathrm w}$ correspondingly. Points in the current configuration are also referred to by their physical coordinates $(x_t,y_t,z_t)^\top$. Following Quarteroni and Formaggia~\cite[p.~73, Sec.~18]{quarteroni2004mathematical}, we introduce the ALE mapping defined as
\begin{equation}
  \mathcal{A}_t : \Omega_0 \to \Omega_t, \qquad
  \boldsymbol{x}_0 \to \boldsymbol{x}_t
  = \boldsymbol{x}_0 + \boldsymbol{d}(\boldsymbol{x}_0,t) = \mathcal{A}_t(\boldsymbol{x}_0),
  \label{eq:ale}
\end{equation}
where $\boldsymbol{d}(\boldsymbol{x}_0,t)$ describes displacement on
the reference domain. We take the reference configuration to be
the configuration at $t=0$, so that $\boldsymbol d(\cdot,0)=\boldsymbol 0$
and $\mathcal A_0$ is the identity. The corresponding
domain velocity on $\Omega_t$ is
$\boldsymbol{w}(\boldsymbol{x}_t,t)
=\partial_t\boldsymbol{d}(\mathcal A_t^{-1}(\boldsymbol{x}_t),t)$.
For space--time integrals and sampling, we write
$\Omega_T:=\{(\boldsymbol{x},t):\boldsymbol{x}\in\Omega_t,\,
0\le t\le T\}$ and
$\Gamma_T^{\mathrm j}:=\{(\boldsymbol{x},t):
\boldsymbol{x}\in\Gamma_t^{\mathrm j},\,0\le t\le T\}$
for $\mathrm j\in\{\mathrm i,\mathrm o,\mathrm w\}$.

The modeled velocity field needs to satisfy the unsteady incompressible Navier--Stokes equations in the domain $\Omega_t$ as in \cite{quarteroni2004mathematical,crosetto2011parallel}. Specifically, the velocity $\boldsymbol{u}$ associated with a pressure $p$ satisfies the system
\begin{equation}
  \begin{cases}
    \dfrac{\partial }{\partial t}\boldsymbol{u} + (\boldsymbol{u} \cdot\nabla)\boldsymbol{u}
      - \nabla\cdot\boldsymbol{\sigma}_{\mathrm f} = \boldsymbol{0}
      & \text{in } \Omega_t,\quad t\in[0,T], \\[4pt]
    \nabla\cdot\boldsymbol{u} = 0
      & \text{in } \Omega_t,\quad t\in[0,T], \\[2pt]
    \boldsymbol{u} = \boldsymbol{u}_0
      & \text{in } \Omega_0,\quad t=0, \\[2pt]
    \boldsymbol{u} = \boldsymbol{g}
      & \text{on } \Gamma_t^{\mathrm i},\quad t\in[0,T], \\[2pt]
    \boldsymbol{\sigma}_{\mathrm f}\boldsymbol{n}_{\mathrm o} = \boldsymbol{0}
      & \text{on } \Gamma_t^{\mathrm o},\quad t\in[0,T], \\[2pt]
    \boldsymbol{u} = \boldsymbol{w}
      & \text{on } \Gamma_t^{\mathrm w},\quad t\in[0,T],
  \end{cases}
  \label{eq:ns-ale}
\end{equation}
for each domain $\Omega_t \subset D$. The first two equations are the fluid momentum equation and incompressibility constraints, respectively; the inlet, outlet, and wall conditions apply on their respective boundary portions. Here $\nu$ denotes the kinematic viscosity, $\boldsymbol{n}_{\mathrm o}$ is the outward unit normal on the outlet, and $\boldsymbol{\sigma}_{\mathrm f} = -p\boldsymbol{I} + 2\nu\boldsymbol{\epsilon}_{\mathrm f}$ is the density-normalized stress tensor for the incompressible fluid, where $\boldsymbol{\epsilon}_{\mathrm f} = \tfrac12(\nabla\boldsymbol{u}+\nabla\boldsymbol{u}^\top)$. Throughout, pressure and both fluid and solid stresses are normalized by the fluid density $\rho_{\mathrm f}$. System~\eqref{eq:ns-ale} is written in Eulerian form on the moving domain: the time derivative is taken at a fixed current point $\boldsymbol x_t$. The equivalent ALE form takes the time derivative at a fixed reference point $\boldsymbol x_0$ and replaces the convective velocity by $\boldsymbol u-\boldsymbol w$, that is, $\partial_t\boldsymbol u|_{\boldsymbol x_0}+((\boldsymbol u-\boldsymbol w)\cdot\nabla)\boldsymbol u$. We use the Eulerian form because the velocity network takes the current coordinates $(\boldsymbol x_t,t)$ as input (Section~\ref{sec:num-arch}); the ALE map enters through the moving domain and the wall velocity $\boldsymbol w$.

Volumetric FSI models~\cite{crosetto2011parallel} impose the linear-elastic momentum equation in the solid wall and traction continuity at the interface. Here we represent only the wall interface and ignore the solid volume; the coupled momentum method~\cite{figueroa2006coupled} similarly represents the vessel wall as a thin membrane on the fluid boundary. We therefore use traction continuity and the
interface trace of the solid momentum operator to describe the wall interface model:
\begin{equation}
  \begin{aligned}
  J_t(\boldsymbol{\sigma}_{\mathrm f}\circ\mathcal{A}_t)
    \boldsymbol F_t^{-\top}\boldsymbol{n}_{\mathrm w}
    &= \boldsymbol{\sigma}_{\mathrm s}\boldsymbol{n}_{\mathrm w},\\
  \frac{\rho_{\mathrm s}}{\rho_{\mathrm f}}
    \frac{\partial^2\boldsymbol d}{\partial t^2}
    -\nabla_{\boldsymbol x_0}\cdot\boldsymbol{\sigma}_{\mathrm s}
    &=\boldsymbol 0,
  \end{aligned}
  \qquad \text{on } \Gamma_0^{\mathrm w}.
  \label{eq:interface}
\end{equation}
Here $\nabla_{\boldsymbol x_0}$ denotes the spatial gradient with
respect to the reference coordinate $\boldsymbol x_0$, with
$(\nabla_{\boldsymbol x_0}\boldsymbol d)_{ij}=\partial d_i/\partial x_{0,j}$.
In~\eqref{eq:interface},
$\boldsymbol F_t:=\nabla_{\boldsymbol x_0}\mathcal A_t
=\boldsymbol I+\nabla_{\boldsymbol x_0}\boldsymbol d$ is the deformation
gradient and $J_t:=\det\boldsymbol F_t$. By Nanson's formula,
$J_t\boldsymbol F_t^{-\top}\boldsymbol n_{\mathrm w}$ is the
current normal-area vector per unit reference area. The structure stress tensor
$\boldsymbol{\sigma}_{\mathrm s} = \lambda_{\mathrm s}\,\mathrm{tr}(\boldsymbol{\epsilon}_{\mathrm s})\,\boldsymbol{I}
  + 2\mu_{\mathrm s}\,\boldsymbol{\epsilon}_{\mathrm s}$
is the linearized nominal stress of the vessel wall in reference
coordinates, with
$\boldsymbol{\epsilon}_{\mathrm s}
=\tfrac12\big(\nabla_{\boldsymbol x_0}\boldsymbol{d}
+(\nabla_{\boldsymbol x_0}\boldsymbol{d})^\top\big)$,
and $\lambda_{\mathrm s},\mu_{\mathrm s}$ are the Lam\'e parameters
describing the elastic response of the vessel wall. The ratio
$\rho_{\mathrm s}/\rho_{\mathrm f}$ appears because stresses have
been divided by the fluid density. The solid momentum operator is evaluated
on $\Gamma_0^{\mathrm w}$. In
Equation~\eqref{eq:interface}, $\boldsymbol n_{\mathrm w}$ is the
outward unit normal to the reference fluid domain $\Omega_0$ on
$\Gamma_0^{\mathrm w}$ at
$\boldsymbol x_0$. The fluid stress $\boldsymbol\sigma_{\mathrm f}$
is evaluated at $\boldsymbol x_t=\mathcal A_t(\boldsymbol x_0)$,
while $J_t$, $\boldsymbol\sigma_{\mathrm s}$, and $\boldsymbol F_t$
are evaluated at $\boldsymbol x_0$.

\paragraph{Unknowns and known quantities}
In our problem setting, the noisy data $\tilde{\boldsymbol{u}}$ is available from imaging data. The true velocity field $\boldsymbol{u}$, its initial condition $\boldsymbol{u}_0$, the displacement function $\boldsymbol{d}$, the inlet trace $\boldsymbol{g}=\boldsymbol{u}|_{\Gamma_t^{\mathrm i}}$, and the reference domain $\Omega_0$ are unknown. In the numerical method, both the initial condition and the inlet trace are obtained from the learned velocity field rather than represented by separate networks. Also, we treat the fluid's kinematic viscosity $\nu$, wall density $\rho_{\mathrm s}$, and Lam\'e parameters $(\lambda_{\mathrm s},\mu_{\mathrm s})$ as prescribed physical parameters, following~\cite{zhang2026coupled,aguayo2023distributed}. 

Our goal is to reconstruct $(\boldsymbol{u},\boldsymbol{d},\Omega_0)$, together with the inflow condition $\boldsymbol{g}$ and initial condition $\boldsymbol{u}_0$, such that the modeled velocity field is consistent with the observed data and with the fluid and vessel wall mechanics. To quantify the data misfit, we define a discrepancy functional measuring the $L^2$ distance between the measured and simulated velocities as
\begin{equation}
  \mathcal L_{\mathrm{data}}(\boldsymbol{u})
  = \|\tilde{\boldsymbol{u}} - \boldsymbol{u}\|^2_{L^2(\Omega_T)}.
  \label{eq:data-misfit}
\end{equation}
Thus, the fluid data discrepancy is measured within the moving fluid
domain over the observation interval.

Starting from an initial guess $\Omega_{\mathrm{ref}}$, for a reference-domain correction $\boldsymbol{\zeta}$ and time-dependent
displacement $\boldsymbol{d}$, we aim to obtain accurate reference domain
$\Omega_0=(\boldsymbol{I}+\boldsymbol{\zeta})(\Omega_{\mathrm{ref}})$ and time-dependent current frames
$\Omega_t=(\boldsymbol{I}+\boldsymbol{d}(\cdot,t))(\Omega_0)$.  With this geometry
expressed by $\boldsymbol{\zeta}$ and $\boldsymbol{d}$, let $\mathcal U_{\mathrm{fluid}}$ map the inlet and initial data
$(\boldsymbol{g},\boldsymbol{u}_0)$ to
the solution of the fluid system~\eqref{eq:ns-ale}.  The inverse problem is
then posed over $(\boldsymbol{g},\boldsymbol{u}_0,\boldsymbol{\zeta},\boldsymbol{d})$ as
\begin{equation}
  \begin{aligned}
  \min_{\boldsymbol{g},\,\boldsymbol{u}_0,\,\boldsymbol{\zeta},\,\boldsymbol{d}}\quad
  &\mathcal L_{\mathrm{data}}\big(
    \boldsymbol{u}=\mathcal U_{\mathrm{fluid}}
    [\boldsymbol{g},\boldsymbol{u}_0;\boldsymbol{\zeta},\boldsymbol{d},\nu]\big)\\
  &+ \text{geometry regularization}\\
  &+ \alpha_{\mathrm{str}}\mathcal L_{\mathrm{str}}(\boldsymbol{d})
  \end{aligned}
  \label{eq:total-opt}
\end{equation}
subject to the fluid constraint~\eqref{eq:ns-ale}, with both
interface relations in~\eqref{eq:interface} enforced through
$\mathcal L_{\mathrm{str}}$.  Sections~\ref{sec:fluid}--\ref{sec:structure}
make the three terms precise.

However, the problem is complex with multiple physics interacting.
Thus, we split the total problem into subproblems. Since the optimization
problem involves solutions affected by the computational domain, the
first step is to isolate the optimization on factors defining the
computational domain, namely $\boldsymbol{d}$ and $\Omega_0$. The remaining
part forms the Navier--Stokes system on a prescribed moving domain,
which is our fluid subproblem. The isolated domain
problem can then be solved through a registration approach, where the
registration mapping mimics the domain gradient; this is our geometry
subproblem. Besides solving for an accurate domain, the elastic response of the wall is
governed by a known constitutive law that links the reconstructed wall motion to the fluid loading it experiences. We refer to this regularization mechanism as the structure regularization.

\begin{figure}
    \centering
    \includegraphics[width=\linewidth]{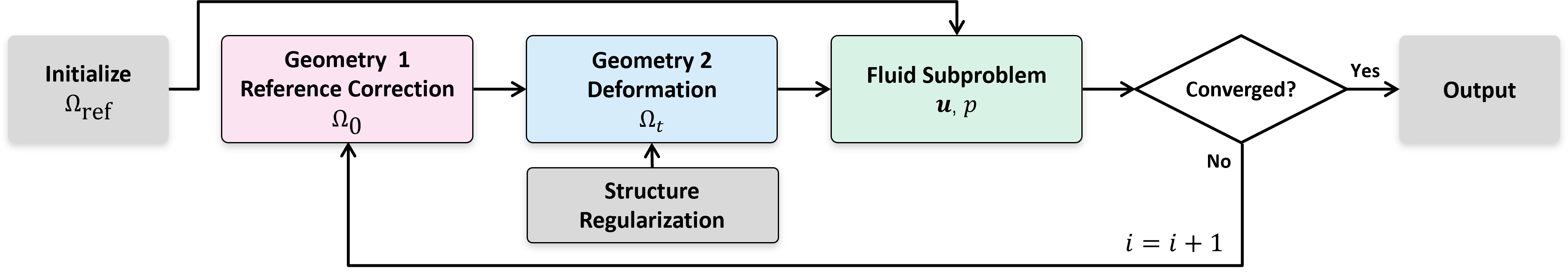}
    \caption{Gauss--Seidel-type alternating reconstruction pipeline. An initial fluid
    subproblem is solved on the supplied prior mask $\Omega_0^0$ with
    zero displacement, before any geometry update. Geometry~1 then
    corrects the reference domain $\Omega_0$. Geometry~2
    estimates the time-dependent domain $\Omega_t$ using registration
    and structure regularization. The fluid subproblem then updates
    velocity and pressure on the resulting domain. After each fluid
    update, the stopping test either returns the reconstruction or
    begins another geometry update.}
    \label{fig:pipeline}
\end{figure}

\subsection{Fluid Problem}
\label{sec:fluid}

For the fluid subproblem, we assume that the domain $\Omega_t$,
displacement $\boldsymbol{d}$, and kinematic viscosity $\nu$ are fixed and
known. With the geometry held fixed, the fluid update minimizes the
discrepancy between the observed velocity field
$\tilde{\boldsymbol{u}}$ and the modeled solution $\boldsymbol{u}$, which
satisfies the Navier--Stokes equations~\eqref{eq:ns-ale} on the moving
domain. In this setting, we redefine the solution
operator $\mathcal U_{\mathrm{fluid}} : (\boldsymbol{g},\boldsymbol{u}_0)\mapsto\boldsymbol{u}$ that
maps candidate inlet and initial conditions $(\boldsymbol{g},\boldsymbol{u}_0)$ to
the corresponding velocity field $\boldsymbol{u}$, while incorporating fixed domain $\Omega_t$.

The resulting optimization problem is a partial minimization over
$(\boldsymbol{g},\boldsymbol{u}_0)$ with
$\boldsymbol{u} = \mathcal{U}_{\mathrm{fluid}}[\boldsymbol{g},\boldsymbol{u}_0]$:
\begin{equation}
  \min_{\boldsymbol{g},\,\boldsymbol{u}_0}
  \mathcal L_{\mathrm{fluid}}
  = \|\tilde{\boldsymbol{u}}-\boldsymbol{u}\|^2_{L^2(\Omega_T)},
  \label{eq:fluid-obj}
\end{equation}
subject to the model constraint~\eqref{eq:ns-ale} with known
$\boldsymbol{d}$, computational domain $\Omega_0$.

\subsection{Geometry Problem}
\label{sec:geometry}

The geometry subproblem has two steps: a correction of the reference
domain followed by a sequential registration of velocity fields.
Starting from the supplied prior mask $\Omega_0^0$ and
$\boldsymbol d^0=\boldsymbol 0$, an initial fluid solve gives
$(\hat{\boldsymbol u}^{\,0},\hat p^{\,0})$. At alternation
$n\geq1$, both geometry steps use the preceding fluid estimate
$(\hat{\boldsymbol u}^{\,n-1},\hat p^{\,n-1})$. Step~1 corrects
$\Omega_{\mathrm{ref}}^n:=\Omega_0^{n-1}$ to obtain $\Omega_0^n$.
Step~2 uses this new domain
and the preceding displacement $\boldsymbol d^{n-1}$ to estimate
$\boldsymbol d^n$. The next fluid update then gives
$(\hat{\boldsymbol u}^{\,n},\hat p^{\,n})$.
For either geometry map $m\in\{\mathcal F_0^n,\mathcal A_t^n\}$, let
$Q_{\mathcal F_0^n}=\Omega_{\mathrm{ref}}^n$ and
$Q_{\mathcal A_t^n}=\Omega_0^n\times[0,T]$ be its source set,
and let $\widehat Q_{\mathcal F_0^n}=D$ and
$\widehat Q_{\mathcal A_t^n}=D\times[0,T]$ be the corresponding image-domain sets.
We use a hyperelastic regularizer adapted
from~\cite{burger2013hyperelastic} and a Laplacian regularizer,
\begin{equation}
  \begin{aligned}
  \mathcal L_{\mathrm{hyper}}(m)
    &:=\big\|\phi_l(\nabla m)+\phi_a(\nabla m)
      +\phi_v(\nabla m)\big\|^2_{L^2(\widehat Q_m)},\\
  \mathcal L_{\mathrm{lap}}(m)
    &:=\|\Delta m\|^2_{L^2(Q_m)},
  \end{aligned}
  \label{eq:geometry-energies}
\end{equation}
where, for $X\in\mathbb R^{3\times3}$ with $\det X\ne0$, the penalty
densities are
\begin{equation}
  \begin{aligned}
  \phi_l(X)&:=\tfrac12\|X-\boldsymbol I\|_{\mathrm F}^2,\\
  \phi_a(X)&:=\tfrac12\big(\|\operatorname{cof}X\|_{\mathrm F}^2-3\big)^2,\\
  \phi_v(X)&:=\left(\frac{(\det X-1)^2}{\det X}\right)^2.
  \end{aligned}
  \label{eq:hyperelastic-penalties}
\end{equation}
Here $\|\cdot\|_{\mathrm F}$ is the Frobenius norm. The three terms
penalize length, area, and volume distortion, respectively. In
particular, $\phi_a$ acts on the cofactor of the deformation
gradient, while $\phi_v$ penalizes volume collapse as
$\det\nabla m\to0$. The Laplacian penalty encourages the displacement
in the reference fluid domain to behave as a harmonic extension of

\paragraph{Step 1: Reference-domain estimation}
The preceding initial domain serves as the reference domain
$\Omega_{\mathrm{ref}}^n=\Omega_0^{n-1}$. We estimate a static
correction $\mathcal F_0^n=\boldsymbol I+\boldsymbol\zeta^n$ and
update the initial domain by
\begin{equation}
  \Omega_0^n = \mathcal F_0^n(\Omega_{\mathrm{ref}}^n).
  \label{eq:ref-domain-update}
\end{equation}
The correction $\boldsymbol{\zeta}^n$ is obtained by registering the
initial flow velocity field while regularizing the map:
\begin{equation}
  \begin{aligned}
  \min_{\boldsymbol{\zeta}^n}\quad
  \mathcal L_{\mathrm{geo},0}
    &:=\alpha_{\mathrm{reg},0}\mathcal L_{\mathrm{reg},0}
      +\alpha_{\mathrm{hyper},0}\mathcal L_{\mathrm{hyper}}(\mathcal F_0^n)
      +\alpha_{\mathrm{lap},0}\mathcal L_{\mathrm{lap}}(\mathcal F_0^n),\\
  \mathcal L_{\mathrm{reg},0}
    &:=\big\|\tilde{\boldsymbol{u}}(\cdot,0)
      -\hat{\boldsymbol{u}}^{\,n-1}(\cdot,0)\circ(\mathcal F_0^n)^{-1}
      \big\|^2_{L^2(D)}.
  \end{aligned}
  \label{eq:step1-loss}
\end{equation}
Here $\hat{\boldsymbol{u}}^{\,n-1}(\cdot,0)\circ(\mathcal F_0^n)^{-1}$
is supported on $\Omega_0^n$ and is extended by zero outside it.

\paragraph{Step 2: Time-dependent deformation estimation}
After Step~1 updates the reference domain to $\Omega_0^n$,
Geometry Step~2 expresses the preceding velocity field on this new
reference domain. For $\boldsymbol y\in\Omega_0^n$,
$(\mathcal F_0^n)^{-1}(\boldsymbol y)$ lies in $\Omega_0^{n-1}$,
and $\mathcal A_t^{n-1}\big((\mathcal F_0^n)^{-1}(\boldsymbol y)\big)$
lies in the preceding physical domain $\Omega_t^{n-1}$. We therefore
define the transported velocity image by
\begin{equation}
  \hat{\boldsymbol u}_{\mathrm{img}}^{\,n}(\boldsymbol y,t)
  :=\hat{\boldsymbol u}^{\,n-1}\!\left(
      \mathcal A_t^{n-1}\!\left(
        (\mathcal F_0^n)^{-1}(\boldsymbol y)\right),t\right),
  \qquad \boldsymbol y\in\Omega_0^n.
  \label{eq:reference-corrected-velocity}
\end{equation}
Here $\mathcal A_t^{n-1}:\Omega_0^{n-1}\to\Omega_t^{n-1}$, and
$\hat{\boldsymbol u}^{\,n-1}(\cdot,t)$ is evaluated only on
$\Omega_t^{n-1}$. Thus $\hat{\boldsymbol u}_{\mathrm{img}}^{\,n}$
is supported on $\Omega_0^n$. Its velocity components are transported
as image channels, without changing their vector basis. This image
uses the preceding fluid field and ALE map from the last alternation, both held fixed during
Geometry Step~2.

With $\Omega_0^n$ fixed, we estimate a new ALE map
$\mathcal A_t^n=\boldsymbol I+\boldsymbol d^n(\cdot,t)$, whose image
$\Omega_t^n=\mathcal A_t^n(\Omega_0^n)$ approximates the physical
domain. After extending $\hat{\boldsymbol u}_{\mathrm{img}}^{\,n}$ by zero outside
$\Omega_0^n$, we register
$\hat{\boldsymbol u}_{\mathrm{img}}^{\,n}\circ(\mathcal A_t^n)^{-1}$
with the observed data. The objective also includes geometric regularization
and the structure regularization introduced in
Section~\ref{sec:structure}:
\begin{equation}
  \begin{aligned}
  \min_{\boldsymbol d^n}\quad
  \mathcal L_{\mathrm{geo},t}
    &:=\alpha_{\mathrm{reg}}\mathcal L_{\mathrm{reg},t}
      +\alpha_{\mathrm{hyper}}\mathcal L_{\mathrm{hyper}}(\mathcal A_t^n)
      +\alpha_{\mathrm{lap}}\mathcal L_{\mathrm{lap}}(\mathcal A_t^n)\\
    &\phantom{{}:=}+\alpha_{\mathrm{str}}
      \mathcal L_{\mathrm{str}}(\boldsymbol d^n)
      +\alpha_{\mathrm{init},d}\mathcal L_{\mathrm{init},d},\\
  \mathcal L_{\mathrm{reg},t}
    &:=\big\|\tilde{\boldsymbol{u}}
      -\hat{\boldsymbol u}_{\mathrm{img}}^{\,n}\circ(\mathcal A_t^n)^{-1}
      \big\|^2_{L^2(D\times[0,T])},\\
  \mathcal L_{\mathrm{init},d}
    &:=\|\boldsymbol d^n(\cdot,0)\|^2_{L^2(D)}.
  \end{aligned}
  \label{eq:step2-loss}
\end{equation}
In $\mathcal L_{\mathrm{reg},t}$, the candidate $\mathcal A_t^n$
maps $\Omega_0^n$ to $\Omega_t^n$; hence
$\hat{\boldsymbol u}_{\mathrm{img}}^{\,n}\circ(\mathcal A_t^n)^{-1}$
is supported on $\Omega_t^n$ and can be compared with the observed
velocity there. The structure term uses the preceding fluid
estimate $(\hat{\boldsymbol u}^{\,n-1},\hat p^{\,n-1})$, held fixed
during this geometry update. The subsequent fluid update produces
$\hat{\boldsymbol u}^{\,n}$ on $\Omega_t^n$.
Here $\mathcal L_{\mathrm{hyper}}$ and $\mathcal L_{\mathrm{lap}}$
regularize the geometry, whereas $\mathcal L_{\mathrm{str}}$ favors
displacements consistent with both fluid--wall traction continuity and
the interface-evaluated solid momentum equation. The term
$\mathcal L_{\mathrm{init},d}$ penalizes initial displacement on the
image domain $D$. All $\alpha_{\mathrm{reg},0}$,
$\alpha_{\mathrm{hyper},0}$, $\alpha_{\mathrm{lap},0}$,
$\alpha_{\mathrm{reg}}$, $\alpha_{\mathrm{hyper}}$,
$\alpha_{\mathrm{lap}}$, $\alpha_{\mathrm{str}}$, and
$\alpha_{\mathrm{init},d}$ are nonnegative
loss weights.

Notice that an accurately reconstructed lumen does not uniquely determine pointwise wall displacement, which is also the reason for having the structure regularization. Lemma~\ref{lem:wall-correspondence} distinguishes the deformed wall
as a geometric set from the correspondence of its material points.
Matching the wall shape alone cannot distinguish $b_t$ from
$\tilde b_t^{\varepsilon}$, although they assign different
displacements to $\boldsymbol x_0$. Thus equal domain Dice scores
need not imply equal wall-displacement errors. The proof is in
Appendix~\ref{app:proof-wall}.

\begin{lemma}[Wall displacement is not determined by lumen geometry]
\label{lem:wall-correspondence}
Take a nonempty smooth, relatively open patch
$S\Subset\Gamma_0^{\mathrm w}$ such that
$\overline S\cap(\overline{\Gamma_0^{\mathrm i}}
\cup\overline{\Gamma_0^{\mathrm o}})=\varnothing$.
Let $b_t:=\mathcal A_t|_{\Gamma_0^{\mathrm w}}:
\Gamma_0^{\mathrm w}\to\Gamma_t^{\mathrm w}$ be a diffeomorphism for
each $t\in[0,T]$. Then there exist $\varepsilon_0>0$,
$t_*\in(0,T]$, and $\boldsymbol x_0\in S$, together with a family
of diffeomorphisms
$\tilde b_t^{\varepsilon}:\Gamma_0^{\mathrm w}
\to\Gamma_t^{\mathrm w}$, $|\varepsilon|<\varepsilon_0$, such that
\begin{align*}
  \tilde b_0^{\varepsilon}&=b_0,
  &\tilde b_t^{\varepsilon}(\Gamma_0^{\mathrm w})
    &=b_t(\Gamma_0^{\mathrm w})=\Gamma_t^{\mathrm w}
      &&(t\in[0,T]),\\
  \tilde b_t^{\varepsilon}|_{\Gamma_0^{\mathrm w}\setminus S}
    &=b_t|_{\Gamma_0^{\mathrm w}\setminus S},
  &\tilde b_{t_*}^{\varepsilon}(\boldsymbol x_0)
    &\ne b_{t_*}(\boldsymbol x_0)
      &&(0<|\varepsilon|<\varepsilon_0).
\end{align*}
Moreover, $\tilde b_{t_*}^{\varepsilon_1}(\boldsymbol x_0)
\ne\tilde b_{t_*}^{\varepsilon_2}(\boldsymbol x_0)$ whenever
$\varepsilon_1\ne\varepsilon_2$ are sufficiently small. Thus the
same observed geometry admits infinitely many different pointwise wall displacements $\tilde b_t^{\varepsilon}(\boldsymbol x_0)
-\boldsymbol x_0$.
\end{lemma}


\subsection{Structure Regularization}
\label{sec:structure}

Given the fixed kinematic viscosity $\nu$, the fixed Lam\'e parameters
$(\lambda_{\mathrm s},\mu_{\mathrm s})$, and the velocity/pressure
fields $(\hat{\boldsymbol{u}},\hat p)$ recovered from the fluid subproblem
(Section~\ref{sec:fluid}), the fluid stress is computed directly as
\begin{equation}
  \boldsymbol{\sigma}_{\mathrm f}(\hat{\boldsymbol{u}},\hat p;\nu)
  = -\hat p\,\boldsymbol{I} + 2\nu\,\boldsymbol{\epsilon}_{\mathrm f}(\hat{\boldsymbol{u}}),
  \qquad
  \boldsymbol{\epsilon}_{\mathrm f}(\hat{\boldsymbol{u}}) = \tfrac12\big(\nabla\hat{\boldsymbol{u}}+\nabla\hat{\boldsymbol{u}}^\top\big).
  \label{eq:fluid-stress}
\end{equation}
Recall the linearized solid strain and stress from
Equation~\eqref{eq:interface},
\begin{equation}
  \boldsymbol{\epsilon}_{\mathrm s}(\boldsymbol{d})
  = \tfrac12\big(\nabla_{\boldsymbol x_0}\boldsymbol{d}
    +(\nabla_{\boldsymbol x_0}\boldsymbol{d})^\top\big),
  \qquad
  \boldsymbol{\sigma}_{\mathrm s}(\boldsymbol{d};\lambda_{\mathrm s},\mu_{\mathrm s})
  = \lambda_{\mathrm s}\,\mathrm{tr}\big(\boldsymbol{\epsilon}_{\mathrm s}(\boldsymbol{d})\big)\,\boldsymbol{I}
  + 2\mu_{\mathrm s}\,\boldsymbol{\epsilon}_{\mathrm s}(\boldsymbol{d}).
  \label{eq:solid-stress}
\end{equation}

Because registration can admit different wall correspondences, we
penalize both relations in~\eqref{eq:interface} for the candidate
displacement $\boldsymbol d$ produced by the geometry subproblem:
\begin{equation}
  \begin{aligned}
  \mathcal L_{\mathrm{str}}(\boldsymbol{d})
  := {}&\Big\|\frac{1}{\lambda_{\mathrm s}}\big(
    \boldsymbol{\sigma}_{\mathrm s}
      (\boldsymbol{d};\lambda_{\mathrm s},\mu_{\mathrm s})
    - J_t (\boldsymbol{\sigma}_{\mathrm f}(\hat{\boldsymbol{u}},\hat p;\nu)
      \circ\mathcal A_t)\boldsymbol F_t^{-\top}
    \big)\boldsymbol{n}_{\mathrm w}
  \Big\|^2_{L^2(\Gamma_0^{\mathrm w}\times[0,T])}\\
  &+\Big\|\frac{1}{\lambda_{\mathrm s}}\left(
    \frac{\rho_{\mathrm s}}{\rho_{\mathrm f}}
      \frac{\partial^2\boldsymbol d}{\partial t^2}
    -\nabla_{\boldsymbol x_0}\cdot
      \boldsymbol\sigma_{\mathrm s}
        (\boldsymbol d;\lambda_{\mathrm s},\mu_{\mathrm s})
  \right)\Big\|^2_{L^2(\Gamma_0^{\mathrm w}\times[0,T])}.
  \end{aligned}
  \label{eq:str-residual}
\end{equation}

The reference-domain factors and $\boldsymbol n_{\mathrm w}$ are
evaluated at $\boldsymbol x_0\in\Gamma_0^{\mathrm w}$; the fluid stress is
evaluated at $\boldsymbol x_t=\mathcal A_t(\boldsymbol x_0)$. Both
terms are normalized by $\lambda_{\mathrm s}$ and enter with the same
weight $\alpha_{\mathrm{str}}$. The second term evaluates the
reference-coordinate solid momentum operator on the wall using the extended
displacement field. 
Minimizing $\mathcal L_{\mathrm{str}}$ jointly with the registration
loss in~\eqref{eq:step2-loss} therefore favors wall motions consistent
with the fluid traction and the interface-evaluated elastic dynamics.

\subsection{Partial Optimality of Alternating Reconstruction}
\label{sec:alternating-limit}

We consider converging sequences of the Gauss--Seidel-type alternating updates in
Figure~\ref{fig:pipeline} and Algorithm~\ref{alg:overall}, as in the
steady two-dimensional analysis~\cite{zhang2026coupled}. To distinguish
the correction made in one iteration from
the map out of the original domain $\Omega_0^0$, set
$\mathcal G_0^0=\boldsymbol I$ and
$\mathcal G_0^n:=\mathcal F_0^n\circ\mathcal G_0^{n-1}$. Then
$\Omega_0^n=\mathcal G_0^n(\Omega_0^0)$, whereas
$\mathcal F_0^n$ maps $\Omega_0^{n-1}$ to $\Omega_0^n$.
For fixed geometry, let
$(\boldsymbol u,p)=\mathcal S_{\mathrm f}
[\boldsymbol g,\boldsymbol u_0;\mathcal G,\boldsymbol d]$
denote a solution of~\eqref{eq:ns-ale}, with velocity component
$\mathcal U_{\mathrm{fluid}}$ as defined above, on the moving domain
$(\boldsymbol I+\boldsymbol d(\cdot,t))\circ
\mathcal G(\Omega_0^0)$. We compare fields on changing domains through
pullbacks to $\Omega_0^0$.
The three subproblem objectives can be written as
\begin{equation}
  \begin{aligned}
  E_0(\boldsymbol\zeta;\boldsymbol u,\mathcal G_{\mathrm{prev}})
    &:=\mathcal L_{\mathrm{geo},0}
      \quad\text{of~\eqref{eq:step1-loss}},\\
  E_t(\boldsymbol d;\boldsymbol u,p,\mathcal G,
      \mathcal F_0,\boldsymbol d_{\mathrm{prev}})
    &:=\mathcal L_{\mathrm{geo},t}
      \quad\text{of~\eqref{eq:step2-loss}},\\
  E_{\mathrm f}(\boldsymbol g,\boldsymbol u_0;
      \mathcal G,\boldsymbol d)
    &:=\mathcal L_{\mathrm{fluid}}
      \quad\text{of~\eqref{eq:fluid-obj}}.
  \end{aligned}
  \label{eq:alternating-stage-losses}
\end{equation}
In $E_0$, the source domain is
$\mathcal G_{\mathrm{prev}}(\Omega_0^0)$, and
$\boldsymbol\zeta$ is the candidate correction from that domain.
In $E_t$ and $E_{\mathrm f}$, the updated initial domain is
$\mathcal G(\Omega_0^0)$. The geometry objectives hold the preceding
fluid solution fixed. In $E_t$, the registered velocity is
$\boldsymbol u_{\mathrm{img}}\circ\mathcal A_t^{-1}$, where
$\boldsymbol u_{\mathrm{img}}=
\boldsymbol u\circ(\boldsymbol I+\boldsymbol d_{\mathrm{prev}})
\circ\mathcal F_0^{-1}$ on the updated initial domain, as in
\eqref{eq:reference-corrected-velocity}. The structure loss uses the
preceding $(\boldsymbol u,p)$ in its traction term.

For the following result, let $\overline D\subset\mathbb R^3$ be the
compact image domain and $[0,T]$ the observation interval. After the
$n$th fluid update, write $\boldsymbol u^n$, $p^n$,
$\boldsymbol g^n$, $\boldsymbol u_0^n$, $\boldsymbol\zeta^n$, and
$\boldsymbol d^n$ for the fluid state, inlet and initial data,
reference correction, and displacement. The fluid state satisfies
\[
  (\boldsymbol u^n,p^n)=\mathcal S_{\mathrm f}
  [\boldsymbol g^n,\boldsymbol u_0^n;
   \mathcal G_0^n,\boldsymbol d^n].
\]
Set $\mathcal F_0^n:=\boldsymbol I+\boldsymbol\zeta^n$,
$\mathcal A_t^n:=\boldsymbol I+\boldsymbol d^n(\cdot,t)$, and
$\mathcal M_t^n:=\mathcal A_t^n\circ\mathcal G_0^n$.
Whenever the inverses exist, define the relative geometry updates by
\begin{equation}
  \mathcal C_0^n:=\mathcal G_0^n\circ(\mathcal G_0^{n-1})^{-1}
  =\mathcal F_0^n,
  \qquad
  \mathcal C_t^n:=\mathcal M_t^n\circ(\mathcal M_t^{n-1})^{-1}.
  \label{eq:relative-geometry-updates}
\end{equation}

We make the following assumptions for the alternating sequence.
\begin{enumerate}[label=(\roman*)]
\item After pullback to fixed reference sets, the fluid and geometry
fields converge in norms controlling the fields and derivatives in
the three objectives. Denote their limits by $\boldsymbol u^*$,
$p^*$, $\boldsymbol g^*$, $\boldsymbol u_0^*$,
$\boldsymbol d^*$, respectively. The maps $\mathcal G_0^n$ and
$\mathcal M_t^n$ extend to
$C^1$ diffeomorphisms of $\overline D$ onto itself.
$\mathcal G_0^n$ converge in $C^1(\overline D)$ to $\mathcal G_0^*$,
and the relative corrections $\boldsymbol\zeta^n$ converge to
$\boldsymbol 0$ in a norm controlling $E_0$. Moreover,
$\mathcal M_t^n$ converge in $C^1(\overline D)$, uniformly in $t$,
to $\mathcal M_t^*=(\boldsymbol I+\boldsymbol d^*(\cdot,t))
\circ\mathcal G_0^*$.
The limiting maps
$\mathcal G_0^*$ and $\mathcal M_t^*$ are also $C^1$
diffeomorphisms of $\overline D$ onto itself for every $t\in[0,T]$.
\item The fluid solution map $\mathcal S_{\mathrm f}$ is
single-valued and continuous near the limiting inlet data, initial
data, and geometry. The three objectives in
\eqref{eq:alternating-stage-losses} are continuous in their displayed
arguments on admissible neighborhoods of the limit, including the
resampling in $E_t$.
\item There are fixed admissible neighborhoods $N_{\mathrm f}$,
$N_0$, and $N_t$ of $(\boldsymbol g^*,\boldsymbol u_0^*)$,
$\boldsymbol 0$, and $\boldsymbol d^*$, respectively, after pullback
to common reference sets, and
nonnegative errors
$\varepsilon_{\mathrm f,n},\varepsilon_{0,n},\varepsilon_{t,n}\to0$,
such that, for all sufficiently large $n$,
\begin{align}
 E_0(\boldsymbol\zeta^n;\boldsymbol u^{n-1},\mathcal G_0^{n-1})
 &\le E_0(\boldsymbol\zeta;\boldsymbol u^{n-1},
              \mathcal G_0^{n-1})+\varepsilon_{0,n}
 \nonumber\\
 E_t(\boldsymbol d^n;\boldsymbol u^{n-1},p^{n-1},
       \mathcal G_0^n,\mathcal F_0^n,\boldsymbol d^{n-1})
 &\le E_t(\boldsymbol d;\boldsymbol u^{n-1},p^{n-1},
           \mathcal G_0^n,\mathcal F_0^n,\boldsymbol d^{n-1})
       +\varepsilon_{t,n}
 \nonumber\\
 E_{\mathrm f}(\boldsymbol g^n,\boldsymbol u_0^n;
              \mathcal G_0^n,\boldsymbol d^n)
 &\le E_{\mathrm f}(\boldsymbol g,\boldsymbol u_0;
                   \mathcal G_0^n,\boldsymbol d^n)
       +\varepsilon_{\mathrm f,n}.
 \label{eq:asymptotic-stage-optimality}
\end{align}
The three inequalities hold for every
$\boldsymbol\zeta\in N_0$, $\boldsymbol d\in N_t$, and
$(\boldsymbol g,\boldsymbol u_0)\in N_{\mathrm f}$, respectively.
\end{enumerate}

\begin{theorem}[Partial optimality of alternating reconstruction]
\label{thm:alternating-limit}
Under assumptions~(i)--(iii), the relative maps $\mathcal C_0^n$ and
$\mathcal C_t^n$ in~\eqref{eq:relative-geometry-updates} converge
uniformly to the identity on $\overline D$, the latter uniformly in
$t\in[0,T]$. Moreover,
$(\boldsymbol u^*,p^*)=\mathcal S_{\mathrm f}
[\boldsymbol g^*,\boldsymbol u_0^*;
 \mathcal G_0^*,\boldsymbol d^*]$, and the limiting fields satisfy
the partial local optimality relations
\begin{equation}
  \begin{aligned}
  E_0(\boldsymbol 0;\boldsymbol u^*,\mathcal G_0^*)&\le
    E_0(\boldsymbol\zeta;\boldsymbol u^*,\mathcal G_0^*),\\
  E_t(\boldsymbol d^*;\boldsymbol u^*,p^*,\mathcal G_0^*,
      \boldsymbol I,\boldsymbol d^*)&\le
    E_t(\boldsymbol d;\boldsymbol u^*,p^*,\mathcal G_0^*,
      \boldsymbol I,\boldsymbol d^*),\\
  E_{\mathrm f}(\boldsymbol g^*,\boldsymbol u_0^*;
       \mathcal G_0^*,\boldsymbol d^*)&\le
    E_{\mathrm f}(\boldsymbol g,\boldsymbol u_0;
       \mathcal G_0^*,\boldsymbol d^*).
  \end{aligned}
  \label{eq:alternating-partial-minima}
\end{equation}
for every $\boldsymbol\zeta\in N_0$, $\boldsymbol d\in N_t$, and
$(\boldsymbol g,\boldsymbol u_0)\in N_{\mathrm f}$, respectively.
\end{theorem}

The first inequality in~\eqref{eq:alternating-partial-minima} concerns a further relative correction of the limiting initial domain $\mathcal G_0^*(\Omega_0^0)$; it does not identify the cumulative map $\mathcal G_0^*$ with the identity. The observed stabilization of the alternating reconstruction motivates the convergence assumed in~(i). Assumption~(ii) is plausible for smooth network fields and regular inverse mappings. Since each stage minimizes its own objective, the approximate minimization assumed in~(iii) is consistent with sufficiently accurate subproblem solves. The proof is given in Appendix~\ref{app:proof-alternating}.

\section{Numerical Method}
\label{sec:numerics}

Section~\ref{sec:num-arch} introduces the network architecture and optimization rule shared by the subproblems of Section~\ref{sec:formulation}. Sections~\ref{sec:num-fluid} and~\ref{sec:num-geometry} specify the fluid and structure-regularized geometry updates, and Algorithm~\ref{alg:overall} assembles the reconstruction scheme.

\subsection{PirateNets Architecture}
\label{sec:num-arch}

We represent the flow and geometry with four types of PirateNet modules~\cite{wang2024piratenets}: one each for velocity, pressure, reference-domain correction, and wall displacement. They share a network template with random Fourier features and adaptive residual blocks but have independent weights.

\paragraph{Random Fourier Feature Embedding}
To alleviate the spectral bias of standard MLPs~\cite{wang2021eigenvector}, an input $\boldsymbol{\xi}\in\mathbb R^{d_{\mathrm{in}}}$ (space, or space--time) is first embedded as
\begin{equation}
  \Phi(\boldsymbol{\xi}) = \big[\cos(\boldsymbol\omega_k\xi_k);\ \sin(\boldsymbol\omega_k\xi_k)\big]_{k=1}^{d_{\mathrm{in}}},
  \qquad \boldsymbol\omega_k\in\mathbb R^{m_k},
  \label{eq:rff}
\end{equation}
where coordinate $\xi_k$ has its own $m_k$ frequencies, initialized with independent $\mathcal N(0,s^2)$ entries with frequency scale $s>0$~\cite{tancik2020fourier}. The embedding is then linearly re-encoded into a pair of feature tensors $U=\varrho(W_1\Phi(\boldsymbol{\xi})+b_1)$, $V=\varrho(W_2\Phi(\boldsymbol{\xi})+b_2)$, where $\varrho$ is the activation function. The frequencies $\boldsymbol\omega_k$, $W_1$, $W_2$, $b_1$, and $b_2$ are learnable.

\paragraph{Adaptive Residual Blocks}
For $l=1,\dots,L$, block $l$ gates its input $\boldsymbol{x}^{(l)}$, with $\boldsymbol{x}^{(1)}:=\Phi(\boldsymbol{\xi})$, through $U,V$ twice,
\begin{equation}
\begin{gathered}
f^{(l)}=\varrho(W_1^{(l)}\boldsymbol{x}^{(l)}+b_1^{(l)}),\quad
  z_1^{(l)}=f^{(l)}\odot U+(1-f^{(l)})\odot V,\\
  g^{(l)}=\varrho(W_2^{(l)}z_1^{(l)}+b_2^{(l)}),\quad
  z_2^{(l)}=g^{(l)}\odot U+(1-g^{(l)})\odot V,\\
  h^{(l)}=\varrho(W_3^{(l)}z_2^{(l)}+b_3^{(l)}),
\end{gathered}
\label{eq:resblock}
\end{equation}
and combines it with a trainable, zero-initialized skip connection,
\begin{equation}
  \boldsymbol{x}^{(l+1)} = \alpha^{(l)}h^{(l)} + (1-\alpha^{(l)})\boldsymbol{x}^{(l)},
  \qquad \alpha^{(l)}\in\mathbb R,
  \label{eq:skip}
\end{equation}
so that the network behaves as a shallow, near-identity map at initialization and deepens gradually during training. A final linear layer $\boldsymbol{x}^{(L+1)}\mapsto W^{(L+1)}\boldsymbol{x}^{(L+1)}$ produces the
output.

\paragraph{Four Task-Specific Network Roles}
The template defines four network roles, each with independent parameters,
\begin{equation}
  \begin{aligned}
  \hat{\boldsymbol{u}}(\boldsymbol{x},t)
    &=\mathcal N_{\boldsymbol{u}}(\boldsymbol{x},t;\boldsymbol{\theta}_{\boldsymbol{u}}),
  &\qquad
  \hat p(\boldsymbol{x},t)
    &=\mathcal N_p(\boldsymbol{x},t;\boldsymbol{\theta}_p),\\
  \hat{\boldsymbol{\zeta}}(\boldsymbol{x}_0)
    &=\mathcal N_{\boldsymbol{\zeta}}(\boldsymbol{x}_0;\boldsymbol{\theta}_{\boldsymbol{\zeta}}),
  &
  \hat{\boldsymbol{d}}(\boldsymbol{x}_0,t)
    &=\mathcal N_{\boldsymbol{d}}(\boldsymbol{x}_0,t;\boldsymbol{\theta}_{\boldsymbol{d}}).
  \end{aligned}
  \label{eq:networks}
\end{equation}
Here $\hat{\boldsymbol{\zeta}}$ is the static reference-domain correction trained in Geometry Step~1, while $\hat{\boldsymbol{d}}$ is the time-dependent deformation trained in Geometry Step~2. The velocity and pressure modules take space--time coordinates $(x,y,z,t)$, with three and one outputs, respectively. The reference-correction module takes spatial reference coordinates $(x_0,y_0,z_0)$, while the wall-displacement module also takes time $t$. The initial condition and inlet velocity are unknown and are represented implicitly by traces of the learned velocity field:
\begin{equation}
  \hat{\boldsymbol{u}}_0(\boldsymbol{x})
  :=\mathcal N_{\boldsymbol{u}}(\boldsymbol{x},0;\boldsymbol{\theta}_{\boldsymbol{u}}),
  \qquad
  \hat{\boldsymbol{g}}(\boldsymbol{x},t)
  :=\mathcal N_{\boldsymbol{u}}(\boldsymbol{x},t;\boldsymbol{\theta}_{\boldsymbol{u}})
  \big|_{\Gamma_t^{\mathrm i}}.
  \label{eq:learned-initial}
\end{equation}
Thus neither a ground-truth initial field nor a prescribed inlet profile is supplied to the fluid subproblem. They are inferred from the data and fluid constraints through the learned velocity field.

The networks are sampled consistently with the ALE mapping~\eqref{eq:ale}. For a reference point $\boldsymbol{x}_0$ drawn from the current binary volume mask $M_{\Omega_0}$,
\begin{equation}
  \boldsymbol{x}_t
  =\mathcal A_t(\boldsymbol{x}_0)
  =\boldsymbol{x}_0+\hat{\boldsymbol{d}}(\boldsymbol{x}_0,t),
  \qquad
  \big(\hat{\boldsymbol{u}}(\boldsymbol{x}_t,t),\hat p(\boldsymbol{x}_t,t)\big)
  =\big(\mathcal N_{\boldsymbol{u}}(\boldsymbol{x}_t,t;
         \boldsymbol{\theta}_{\boldsymbol{u}}),
         \mathcal N_p(\boldsymbol{x}_t,t;\boldsymbol{\theta}_p)\big).
  \label{eq:sample-consistency}
\end{equation}
In the losses below, $\hat{\boldsymbol{u}}$ and $\hat p$ are evaluated at these sampled points unless a boundary set is specified. The estimated domain velocity is the corresponding ALE velocity, $\hat{\boldsymbol{w}}(\boldsymbol{x}_t,t) =\partial_t\hat{\boldsymbol{d}} (\mathcal A_t^{-1}(\boldsymbol{x}_t),t)$. Let $\mathcal D_{\mathrm g}\subset\Omega_T$ denote the interior data and physics collocation set, $\mathcal D_{\mathrm o}\subset \Gamma_T^{\mathrm o}$ the outlet set, $\mathcal D_{\mathrm w}\subset\Gamma_T^{\mathrm w}$ the wall set. For a field $F$ evaluated on any finite set $\mathcal D$, define
\begin{equation}
  \|F\|_{\mathcal D}^2
  :=\frac{1}{|\mathcal D|}
    \sum_{\boldsymbol{z}\in\mathcal D}|F(\boldsymbol{z})|^2.
  \label{eq:l2norm}
\end{equation}
The inverse solver evaluates the networks and physical residuals at sampled coordinates. A reference-domain mesh is used only to generate these points; the displacement network maps them to the current domain. The mesh connectivity does not enter the loss or residual calculations.

Each fluid or geometry update minimizes its weighted sampled loss over the active network parameters using an Adam-type optimizer~\cite{KingBa15}. The losses and fixed quantities are specified below.

\subsection{Fluid Subproblem}
\label{sec:num-fluid}

\begin{figure}[htbp]
    \centering
    \includegraphics[width=\linewidth]{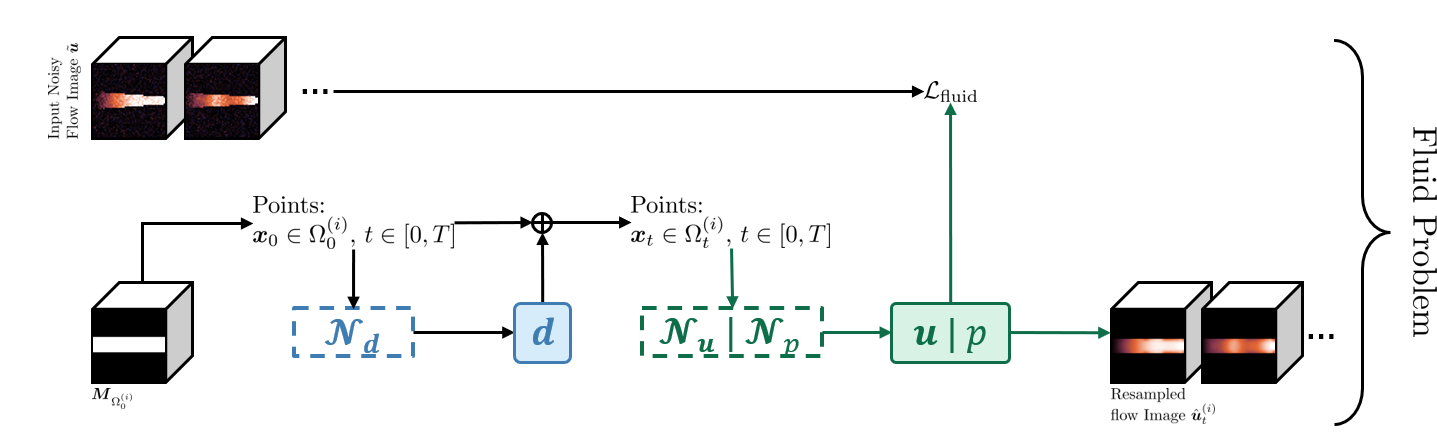}
    \caption{Data flow of the fluid subproblem
    (Section~\ref{sec:num-fluid}). Collocation points $\boldsymbol{x}_0$ drawn
    from the current reference mask $M_{\Omega_0}$ are mapped to the
    current configuration $\boldsymbol{x}_t=\boldsymbol{x}_0+\hat{\boldsymbol{d}}$; the
    velocity/pressure networks are evaluated at $\boldsymbol{x}_t$ to yield
    $(\hat{\boldsymbol{u}},\hat p)$ and the resampled flow image, which together with
    the input noisy image $\tilde{\boldsymbol{u}}$ and the fluid-mechanics
    residuals of Eq.~\eqref{eq:fluid-losses} form $\widehat{\mathcal L}_{\mathrm{fluid}}$.}
    \label{fig:pipeline-fluid}
\end{figure}

The fluid subproblem trains $\boldsymbol{\theta}_{\boldsymbol{u}}$ and
$\boldsymbol{\theta}_p$. Its data, fluid momentum equation, incompressibility,
wall, and outlet losses are
\begin{equation}
  \begin{aligned}
  \mathcal L_{\mathrm{data}} &= \|\tilde{\boldsymbol{u}}-\hat{\boldsymbol{u}}\|^2_{\mathcal D_{\mathrm g}}, \\
  \mathcal L_{\mathrm{mom}} &= \big\|\frac{\partial }{\partial t}\hat{\boldsymbol{u}}+\hat{\boldsymbol{u}}\!\cdot\!\nabla\hat{\boldsymbol{u}}-\nabla\cdot\boldsymbol{\sigma}_{\mathrm f}(\hat{\boldsymbol{u}},\hat p;\nu)\big\|^2_{\mathcal D_{\mathrm g}}, \\
  \mathcal L_{\mathrm{div}} &= \|\nabla\!\cdot\!\hat{\boldsymbol{u}}\|^2_{\mathcal D_{\mathrm g}}, \\
  \mathcal L_{\mathrm{wall}} &= \|\hat{\boldsymbol{u}}-\hat{\boldsymbol{w}}\|^2_{\mathcal D_{\mathrm w}}, \\
  \mathcal L_{\mathrm{out}} &= \|(-\hat p\boldsymbol{I}+2\nu\boldsymbol{\epsilon}_{\mathrm f}(\hat{\boldsymbol{u}}))\boldsymbol{n}_{\mathrm o}\|^2_{\mathcal D_{\mathrm o}},
  \end{aligned}
  \label{eq:fluid-losses}
\end{equation}
using the given $\nu$ and the (fixed, in this subproblem) displacement
$\hat{\boldsymbol{d}}$. The wall velocity $\hat{\boldsymbol{w}}$ is obtained from
the ALE displacement as defined after~\eqref{eq:ale}.
Because $\hat{\boldsymbol{g}}$ is the inlet trace of $\hat{\boldsymbol{u}}$,
an inlet mismatch $\|\hat{\boldsymbol{u}}-\hat{\boldsymbol{g}}\|^2$ would vanish
identically, so no inlet penalty is included. Neither is the initial condition.
The sampled fluid loss is then $\widehat{\mathcal L}_{\mathrm{fluid}} =
\alpha_{\mathrm{data}}\mathcal L_{\mathrm{data}}
+\alpha_{\mathrm{mom}}\mathcal L_{\mathrm{mom}}
+\alpha_{\mathrm{div}}\mathcal L_{\mathrm{div}}
+\alpha_{\mathrm{wall}}\mathcal L_{\mathrm{wall}}
+\alpha_{\mathrm{out}}\mathcal L_{\mathrm{out}}$, minimized over
$(\boldsymbol{\theta}_{\boldsymbol{u}},\boldsymbol{\theta}_p)$. 

Figure~\ref{fig:pipeline-fluid} shows the corresponding data flow:
collocation points sampled from the current reference mask are displaced
to the current configuration by the deformation field $\hat{\boldsymbol{d}}$, the
velocity and pressure networks are evaluated at the deformed points to
produce the resampled flow image, and the discrepancy with the input
noisy image together with the fluid-mechanics residuals is accumulated
into $\widehat{\mathcal L}_{\mathrm{fluid}}$.

\subsection{Structure-Regularized Geometry Problem}
\label{sec:num-geometry}

\begin{figure}[htbp]
    \centering
    \includegraphics[width=\linewidth]{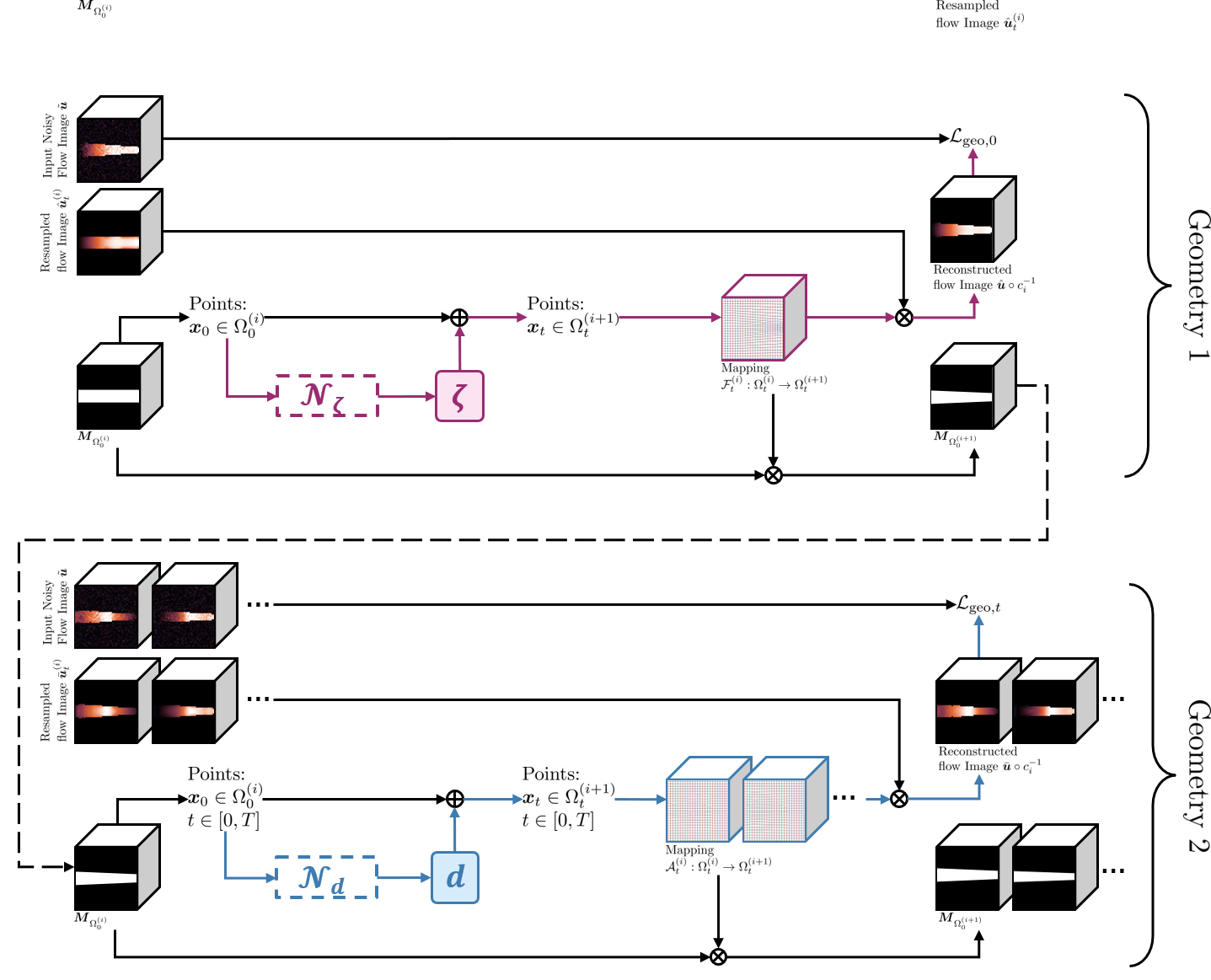}
    \caption{The two geometry updates. Step~1 maps the preceding initial
    domain to the corrected $\Omega_0^n$. Step~2 transports the preceding
    fluid field to $\Omega_0^n$ through
    $\mathcal A_t^{n-1}\circ(\mathcal F_0^n)^{-1}$, then registers the
    resulting image through the new $\mathcal A_t^n$. Both steps use
    hyperelastic and Laplacian regularization; Step~2 also uses
    traction and solid momentum regularization.}
    \label{fig:pipeline-geometry}
\end{figure}

Following the two-step split of Section~\ref{sec:geometry}, the
registration losses $\mathcal L_{\mathrm{reg},0}$ and
$\mathcal L_{\mathrm{reg},t}$ are defined in
Equations~\eqref{eq:step1-loss}--\eqref{eq:step2-loss}. Geometry Step~1
uses the initial-time image $\hat{\boldsymbol u}\circ\mathcal F_0^{-1}$.
Geometry Step~2 forms a time-dependent image
$\hat{\boldsymbol u}_{\mathrm{img}}^{\,n}$ on the corrected $\Omega_0^n$ as in
Equation~\eqref{eq:reference-corrected-velocity}, then registers
$\hat{\boldsymbol u}_{\mathrm{img}}^{\,n}\circ(\mathcal A_t^n)^{-1}$.
The hyperelastic and Laplacian regularizers are those of
Equations~\eqref{eq:geometry-energies}--\eqref{eq:hyperelastic-penalties}.
During training, each $L^2$ norm in these losses is evaluated by the
empirical squared norm in~\eqref{eq:l2norm}. Registration and
hyperelastic points lie on the voxel grid of the image-domain set
$\widehat Q_m$; Laplacian points lie in the corresponding source set
$Q_m$, and structure points on the reference wall. We retain the same loss symbols for these sampled approximations.

\paragraph{Step 1 (reference-domain correction)}
$\mathcal F_0=\boldsymbol{I}+\hat{\boldsymbol{\zeta}}$, trained by
\begin{equation}
  \mathcal L_{\mathrm{geo},0}=\alpha_{\mathrm{reg},0}\mathcal L_{\mathrm{reg},0}
  +\alpha_{\mathrm{hyper},0}\mathcal L_{\mathrm{hyper}}(\mathcal F_0)
  +\alpha_{\mathrm{lap},0}\mathcal L_{\mathrm{lap}}(\mathcal F_0).
  \label{eq:geo-0-loss}
\end{equation}
After this step, $\Omega_0^n\leftarrow
\mathcal F_0^n(\Omega_{\mathrm{ref}}^n)$ as described in Section~\ref{sec:geometry}.

\paragraph{Step 2 (time-dependent deformation)}
For each $\boldsymbol y\in\Omega_0^n$, the preceding velocity network is evaluated
at this point to form the image on the new reference domain:
\[
\hat{\boldsymbol u}_{\mathrm{img}}^{\,n}(\boldsymbol y,t)
:=\mathcal N_{\boldsymbol u}\!\left(
  \mathcal A_t^{n-1}\!\left((\mathcal F_0^n)^{-1}(\boldsymbol y)\right),
  t;\boldsymbol\theta_{\boldsymbol u}^{\,n-1}\right),
\qquad \boldsymbol y\in\Omega_0^n.
\]
This is the previous physical velocity field pulled back to
$\Omega_0^n$. Since $\mathcal A_t^{n-1}\circ(\mathcal F_0^n)^{-1}$ maps
$\Omega_0^n$ into $\Omega_t^{n-1}$, the velocity network is evaluated
only inside $\Omega_t^{n-1}$; outside $\Omega_0^n$ the image is extended
by zero. Then,
we train the new ALE map $\mathcal A_t^n=\boldsymbol I+
\hat{\boldsymbol d}^{\,n}(\cdot,t)$ by
\begin{equation}
  \mathcal L_{\mathrm{geo},t} = \alpha_{\mathrm{reg}}\mathcal L_{\mathrm{reg},t}
  + \alpha_{\mathrm{hyper}}\mathcal L_{\mathrm{hyper}}(\mathcal A_t^n)
  + \alpha_{\mathrm{lap}}\mathcal L_{\mathrm{lap}}(\mathcal A_t^n)
  + \alpha_{\mathrm{str}}\,\widehat{\mathcal L}_{\mathrm{str}}
  + \alpha_{\mathrm{init},d}\,\widehat{\mathcal L}_{\mathrm{init},d}.
  \label{eq:geo-t-loss}
\end{equation}
Here $\widehat{\mathcal L}_{\mathrm{init},d}$ is the sampled counterpart
of $\|\boldsymbol d(\cdot,0)\|^2_{L^2(D)}$ in~\eqref{eq:step2-loss};
it drives the initial displacement towards zero. The structure term evaluates
traction continuity and the wall-evaluated solid momentum equation for the
candidate displacement in Step~2. With $\nu$, $\rho_{\mathrm s}/\rho_{\mathrm f}$,
and $(\lambda_{\mathrm s},\mu_{\mathrm s})$ given and the reconstructed
$(\hat{\boldsymbol{u}},\hat p)$ held constant, the fluid and solid
stresses~\eqref{eq:fluid-stress}--\eqref{eq:solid-stress} determine the
following discrete regularizer at wall collocation points
$\mathcal D_{\mathrm w,0}\subset\Gamma_0^{\mathrm w}\times[0,T]$:
\begin{equation}
  \begin{aligned}
  \widehat{\mathcal L}_{\mathrm{str}}
  = {}&\big\|\frac{1}{\lambda_{\mathrm s}}\big(
    \boldsymbol{\sigma}_{\mathrm s}
      (\hat{\boldsymbol{d}};\lambda_{\mathrm s},\mu_{\mathrm s})
    -J_t\big(\boldsymbol{\sigma}_{\mathrm f}
      (\hat{\boldsymbol{u}},\hat p;\nu)\circ\mathcal A_t\big)
      \boldsymbol F_t^{-\top}
    \big)\boldsymbol{n}_{\mathrm w}
  \big\|^2_{\mathcal D_{\mathrm w,0}}\\
  &+\big\|\frac{1}{\lambda_{\mathrm s}}\left(
    \frac{\rho_{\mathrm s}}{\rho_{\mathrm f}}
      \frac{\partial^2\hat{\boldsymbol d}}{\partial t^2}
    -\nabla_{\boldsymbol x_0}\cdot
      \boldsymbol\sigma_{\mathrm s}
        (\hat{\boldsymbol d};\lambda_{\mathrm s},\mu_{\mathrm s})
    \right)\big\|^2_{\mathcal D_{\mathrm w,0}}.
  \end{aligned}
  \label{eq:str-residual-num}
\end{equation}
Here $\boldsymbol F_t$ and $J_t$ are formed from the candidate
$\mathcal A_t=\boldsymbol I+\hat{\boldsymbol d}$ at the sampled
reference-wall points, where $\boldsymbol n_{\mathrm w}$ is also
evaluated. The fluid stress is evaluated at their current images
$\mathcal A_t(\boldsymbol x_0)$, while the solid momentum term uses
reference-coordinate derivatives of the extended displacement at the
reference-wall samples. This is the sampled form
of~\eqref{eq:str-residual}. During the
Step~2 update, its gradient is propagated through
$\hat{\boldsymbol{d}}$ alone. The fluid-network parameters
$\boldsymbol{\theta}_{\boldsymbol{u}},\boldsymbol{\theta}_p$ and the
physical coefficients $\nu,\lambda_{\mathrm s},\mu_{\mathrm s},
\rho_{\mathrm s}/\rho_{\mathrm f}$ are held
fixed when computing
$\nabla_{\boldsymbol{\theta}_{\boldsymbol{d}}}
\widehat{\mathcal L}_{\mathrm{str}}$. Thus the two structure residuals guide the
geometry update without defining another network or optimization step.

\subsection{Overall Algorithm}
\label{sec:num-algorithm}

Algorithm~\ref{alg:overall} implements Figure~\ref{fig:pipeline}:
after an initial fluid solve, each cycle performs Geometry~1,
Geometry~2, and Fluid, then tests convergence.

\begin{algorithm}[htbp]
\caption{Alternating fluid and structure-regularized geometry updates}
\label{alg:overall}
\begin{algorithmic}[1]
\REQUIRE noisy data $\tilde{\boldsymbol{u}}$; prior reference mask $\Omega_{\mathrm{ref}}$; fixed constants $\nu,\rho_{\mathrm s}/\rho_{\mathrm f},\lambda_{\mathrm s},\mu_{\mathrm s}$
\ENSURE reconstructed $\hat{\boldsymbol{u}},\hat p,\hat{\boldsymbol{u}}_0,
\hat{\boldsymbol{d}},\hat\Omega_0,\hat{\boldsymbol{g}}$
\STATE initialize the four network roles;
       $\Omega_0^0\leftarrow\Omega_{\mathrm{ref}}$,
       $\hat{\boldsymbol d}^{\,0}\leftarrow\boldsymbol 0$,
       $\mathcal A_t^0\leftarrow\boldsymbol I$, $n\leftarrow0$
\STATE \textbf{Fluid} (\S\ref{sec:num-fluid}): train all nine overlapping time windows (\S\ref{sec:exp-details}) on $\Omega_0^0$ to obtain
       $(\hat{\boldsymbol u}^{\,0},\hat p^{\,0})$;
       record $\mathcal L_{\mathrm{data}}^{(0)}$
\STATE $\mathrm{done}\leftarrow\mathrm{false}$ \COMMENT{no preceding data misfit at $n=0$}
\WHILE{$\neg\mathrm{done}$}
  \STATE $n\leftarrow n+1$;
         $\Omega_{\mathrm{ref}}^n\leftarrow\Omega_0^{n-1}$
  \STATE \textbf{Geometry 1} (\S\ref{sec:num-geometry}): use the first time window to minimize $\mathcal L_{\mathrm{geo},0}$ with $\hat{\boldsymbol u}^{\,n-1}$ and obtain $\mathcal F_0^n$;
         $\Omega_0^n\leftarrow\mathcal F_0^n(\Omega_{\mathrm{ref}}^n)$
  \STATE on $\Omega_0^n$, form
         $\hat{\boldsymbol u}_{\mathrm{img}}^{\,n}
         \leftarrow\hat{\boldsymbol u}^{\,n-1}
         \circ\mathcal A_t^{n-1}\circ(\mathcal F_0^n)^{-1}$
         \COMMENT{Eq.~\eqref{eq:reference-corrected-velocity}}
  \STATE \textbf{Geometry 2} (\S\ref{sec:num-geometry}): train all nine time windows to minimize $\mathcal L_{\mathrm{geo},t}$ using
         $\hat{\boldsymbol u}_{\mathrm{img}}^{\,n}\circ(\mathcal A_t^n)^{-1}$,
         $\widehat{\mathcal L}_{\mathrm{str}}$, and
         $\widehat{\mathcal L}_{\mathrm{init},d}$ to obtain
         $\mathcal A_t^n=\boldsymbol I+\hat{\boldsymbol d}^{\,n}$
  \STATE $\Omega_t^n\leftarrow\mathcal A_t^n(\Omega_0^n)$
  \STATE \textbf{Fluid}: train all nine time windows to minimize $\widehat{\mathcal L}_{\mathrm{fluid}}$ on $\Omega_t^n$
         with $\hat{\boldsymbol d}^{\,n}$ fixed to obtain
         $(\hat{\boldsymbol u}^{\,n},\hat p^{\,n})$;
         record $\mathcal L_{\mathrm{data}}^{(n)}$
  \STATE $\mathrm{done}\leftarrow
    \bigl(|\mathcal L_{\mathrm{data}}^{(n)}-\mathcal L_{\mathrm{data}}^{(n-1)}|<\mathrm{tol}\bigr)
    \lor(n=n_{\max})$ \COMMENT{stopping test after Fluid}
\ENDWHILE
\STATE set $\hat{\boldsymbol{u}}_0\leftarrow\hat{\boldsymbol{u}}^{\,n}(\cdot,0)$,
       $\hat p\leftarrow\hat p^{\,n}$,
       $\hat{\boldsymbol{g}}\leftarrow\hat{\boldsymbol{u}}^{\,n}|_{\Gamma_t^{\mathrm i}}$,
       and $\hat\Omega_0\leftarrow\Omega_0^n$
\end{algorithmic}
\end{algorithm}
\FloatBarrier

\section{Experiments}
\label{sec:experiments}

We test the reconstruction method of
Sections~\ref{sec:formulation}--\ref{sec:numerics} on two unsteady
three-dimensional geometries. After the setup in
Section~\ref{sec:exp-details}, we report reconstruction accuracy
(Section~\ref{sec:exp-main}), examine the structure regularizer
(Section~\ref{sec:exp-coupling}), and test sensitivity to misspecified
physical parameters (Section~\ref{sec:exp-sensitivity}). We also
consider a ConvCH case with a Windkessel outlet (Section~\ref{sec:windkessel}).

\subsection{Implementation Details}
\label{sec:exp-details}

\begin{figure}
    \centering
    \includegraphics[width=\linewidth]{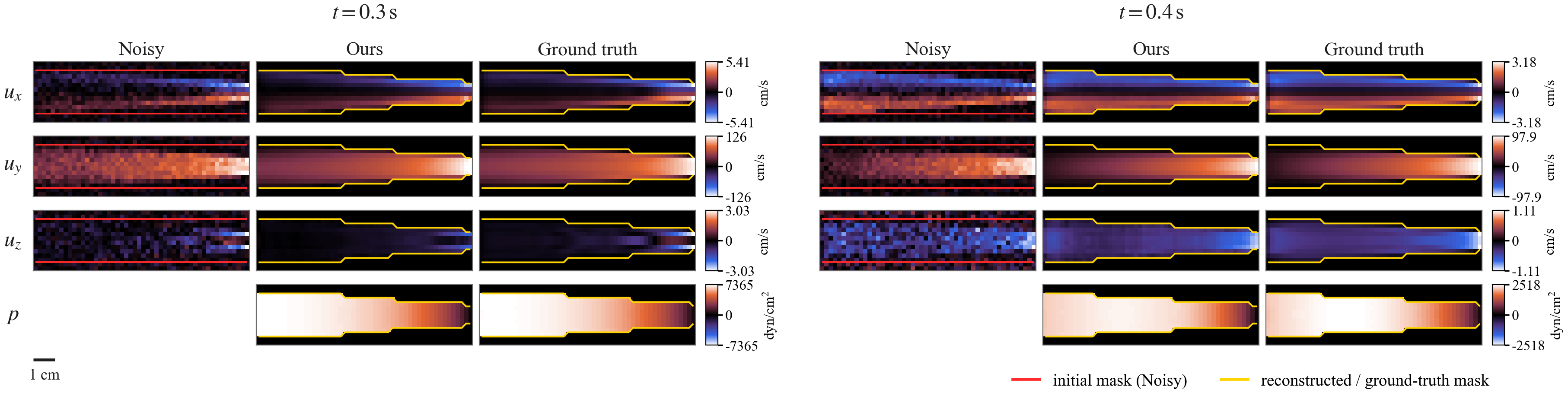}
    \caption{Converging-channel velocity components and pressure at
    $t=0.2$ and $0.3\ \mathrm{s}$. Each time point compares the noisy
    image, reconstruction, and ground truth. Red outlines show the
    initial mask; yellow outlines show the reconstructed and
    ground-truth domains.}
    \label{fig:convCH_slice}
\end{figure}
\begin{figure}
    \centering
    \includegraphics[width=\linewidth]{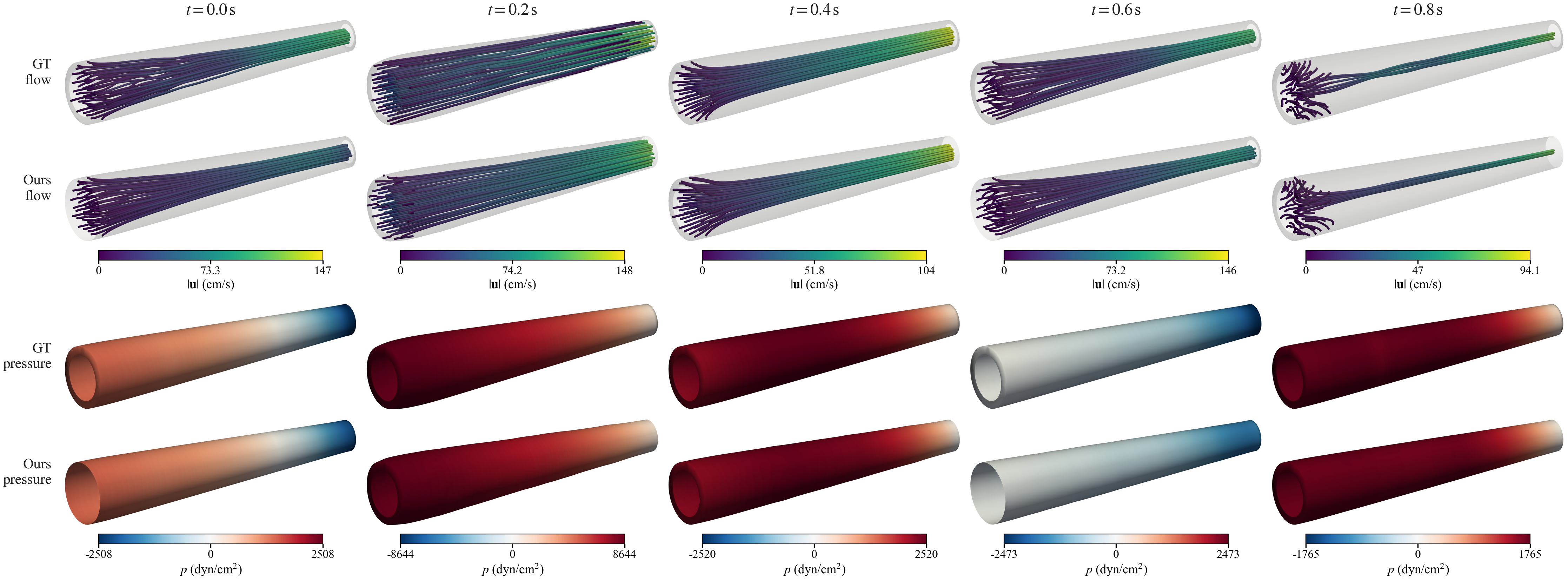}
    \caption{Ground-truth and reconstructed ConvCH flow streamlines
    and wall pressure over the sampled time interval. The pressure color bar range covers the symmetric range, since the plotted pressures include both positive and negative values.}
    \label{fig:convCH_3d}
\end{figure}

\paragraph{Geometry}
The converging channel (ConvCH) is an idealized compliant tube of
length $10\ \mathrm{cm}$, narrowing from $2\ \mathrm{cm}$ at the
inlet to $1\ \mathrm{cm}$ at the outlet. The patient-like compliant
aortic geometry, beginning near the heart, consists of the aortic
arch with its three supra-aortic branches segmented from MRI~\cite{updegrove2017simvascular}.

\paragraph{Physical parameters}
Both geometries use kinematic viscosity $\nu=0.04\ \mathrm{cm^2/s}$, fluid density $\rho_{\mathrm f}=1.06\ \mathrm{g/cm^3}$, and solid density $\rho_{\mathrm s}=1.20\ \mathrm{g/cm^3}$.
The vessel wall has Poisson ratio $\nu_{\mathrm s}=0.3$ and Young's
modulus $E=4\times10^5\ \mathrm{dyn/cm^2}$ for the converging channel
and $E=6\times10^5\ \mathrm{dyn/cm^2}$ for the aorta. The physical
Lam\'e coefficients are
\[
  \lambda_{\mathrm s}^{\mathrm{phys}}
  =\frac{E\nu_{\mathrm s}}{(1+\nu_{\mathrm s})(1-2\nu_{\mathrm s})},
  \qquad
  \mu_{\mathrm s}^{\mathrm{phys}}
  =\frac{E}{2(1+\nu_{\mathrm s})}.
\]
These give $(\lambda_{\mathrm s}^{\mathrm{phys}},
\mu_{\mathrm s}^{\mathrm{phys}})=(2.3077\times10^5,\,
1.5385\times10^5)\ \mathrm{dyn/cm^2}$ for the converging channel and
$(3.4615\times10^5,\,2.3077\times10^5)\ \mathrm{dyn/cm^2}$ for the aorta.
These moduli are lower than typical in vivo arterial values, which
are often of order $10^6$ $\mathrm{dyn/cm^2}$, and $\nu_{\mathrm s}=0.3$
makes the wall compressible; we use them to obtain clearly visible wall
motion in the synthetic tests.
The density-normalized coefficients used in
Equations~\eqref{eq:interface} and~\eqref{eq:str-residual} satisfy
\[
  (\lambda_{\mathrm s},\mu_{\mathrm s})
  =(\lambda_{\mathrm s}^{\mathrm{phys}},
    \mu_{\mathrm s}^{\mathrm{phys}})/\rho_{\mathrm f}.
\]
The reconstruction uses these nominal values except in Section~\ref{sec:exp-sensitivity}, where the assumed coefficients
are perturbed.

\paragraph{Ground truth}
For both geometries, time-dependent FSI simulations in
FEniCS~\cite{scroggs2022basix,bergersen2020turtlefsi}
provide ground-truth velocity $\boldsymbol{u}^{\mathrm{GT}}$,
pressure $p^{\mathrm{GT}}$, and wall displacement
$\boldsymbol{d}^{\mathrm{GT}}$; the deformed mesh defines
$\Omega_t^{\mathrm{GT}}$. Reconstruction uses noisy flow measurements
without supplying $\boldsymbol{u}_0^{\mathrm{GT}}$ or
$\boldsymbol{g}^{\mathrm{GT}}$ to the networks.

\paragraph{Noise model}
We obtain $\boldsymbol{U}^{\mathrm{clean}}$ by interpolating
$\boldsymbol{u}^{\mathrm{GT}}$ and generate
$\boldsymbol{U}^{\mathrm{noisy}}\in
\mathbb R^{11\times3\times50\times50\times50}$ as
\begin{equation}
  \boldsymbol{U}^{\mathrm{noisy}}
  =\boldsymbol{U}^{\mathrm{clean}}
  +0.05\,\max_{x,y,z}\big|\boldsymbol{U}^{\mathrm{clean}}\big|\,
  \boldsymbol{\epsilon},
  \label{eq:noise}
\end{equation}
where the entries of $\boldsymbol\epsilon$ are independent random variables drawn from the standard Gaussian distribution, and the spatial maximum is computed separately at each time and for each velocity component. The sampled observations
$\boldsymbol{U}^{\mathrm{noisy}}$ are denoted by
$\tilde{\boldsymbol{u}}$ in Sections~\ref{sec:formulation}
and~\ref{sec:numerics}.

\paragraph{Hyperparameters and Platform}
Unless varied in an ablation, the loss weights are
$\alpha_{\mathrm{data}}=10^{-1}$,
$\alpha_{\mathrm{mom}}=1$,
$\alpha_{\mathrm{div}}=10^{2}$,
$\alpha_{\mathrm{wall}}=1$,
$\alpha_{\mathrm{out}}=1$,
$\alpha_{\mathrm{reg},0}=\alpha_{\mathrm{reg}}=1$,
$\alpha_{\mathrm{hyper},0}=\alpha_{\mathrm{hyper}}=1$,
$\alpha_{\mathrm{lap},0}=\alpha_{\mathrm{lap}}=10^{2}$,
$\alpha_{\mathrm{str}}=1$, and
$\alpha_{\mathrm{init},d}=1$.
Adam starts at a learning rate of $10^{-3}$, multiplied by $0.8$
after each outer iteration. We use Python~3.9 and PyTorch on an
NVIDIA RTX~4060 Ti.

\paragraph{Network Details}
The velocity and pressure networks each have width $256$; the two
geometry networks each have width $64$. For the comparison in
Table~\ref{tab:main}, the standard PINN baseline uses MLPs with the
same depth and width as the corresponding PirateNet networks.
Each time-dependent network
uses four pairs of temporal random Fourier features. Each spatial
coordinate uses $20$ pairs for velocity and $10$ pairs for pressure
and geometry. Every network has three adaptive residual blocks
($L=3$), each with three hidden transformations. Zero-initialized
final layers in $\mathcal N_{\boldsymbol{\zeta}}$ and
$\mathcal N_{\boldsymbol{d}}$ make the reference correction and wall
displacement vanish initially. Each outer iteration uses nine overlapping
time windows and separate instances of the time-dependent networks.
The fluid networks are trained for $1000$ Adam updates in each of
the nine windows first; Geometry Step~1 then uses $100$ updates in the
first window containing $t=0$ to correct the initial domain; Geometry
Step~2 uses $100$ updates in each of the nine windows to estimate wall
motion. The nine intervals are
$[0,0.2]$, $[0.1,0.3]$, $\ldots$, $[0.6,0.8]$, $[0.7,0.9]$, and
$[0.8,1.0]\ \mathrm{s}$. In regions covered by more than one interval,
the network predictions are combined by distance-weighted interpolation.



\subsection{Main Reconstruction Results}
\label{sec:exp-main}

\begin{figure}
    \centering
    \includegraphics[width=\linewidth]{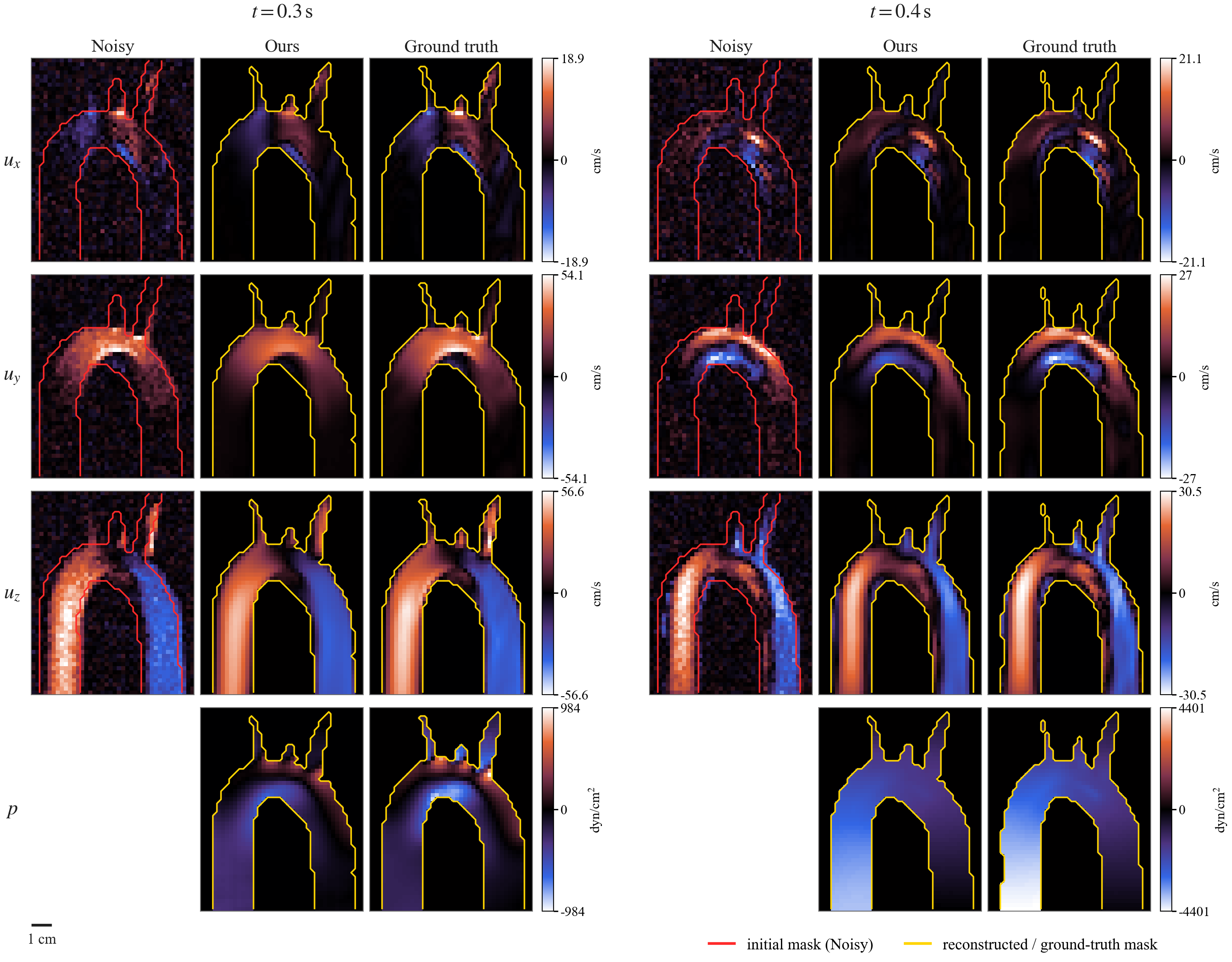}
    \caption{Aortic velocity components and pressure at $t=0.2$ and
    $0.3\ \mathrm{s}$. The noisy image, reconstruction, and ground
    truth are shown side by side, with initial-mask outlines in red
    and reconstructed or ground-truth outlines in yellow.}
    \label{fig:aorta_slice}
\end{figure}
\begin{figure}
    \centering
    \includegraphics[width=\linewidth]{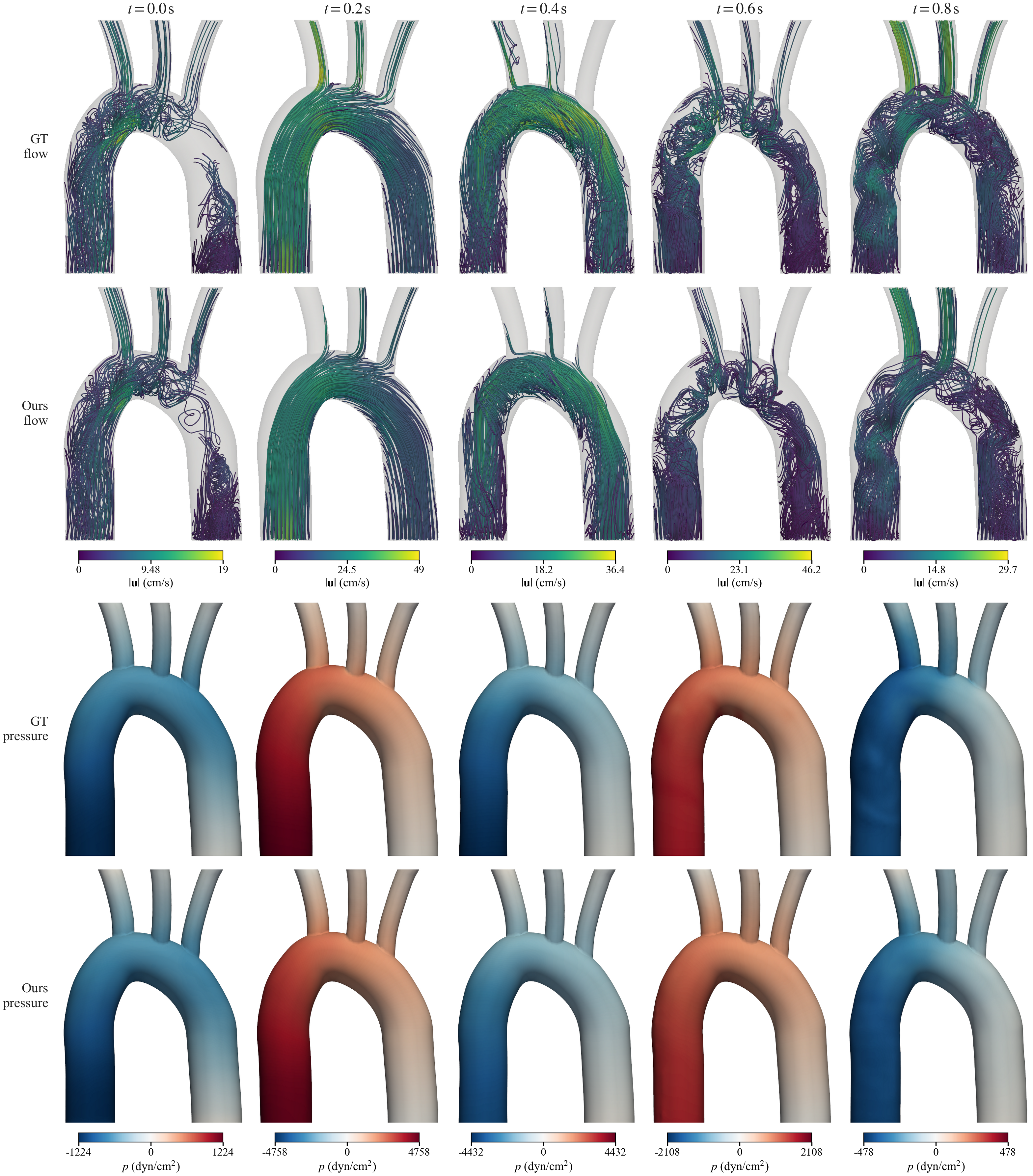}
    \caption{Ground-truth and reconstructed aortic flow streamlines
    and wall pressure at five times from $0$ to $0.8\ \mathrm{s}$. The pressure color bar range covers the symmetric range, since the plotted pressures include both positive and negative values.}
    \label{fig:aorta_3d}
\end{figure}

\begin{table}
\centering
\caption{Reconstruction accuracy for the two test geometries.}
\label{tab:main}
\resizebox{0.9\linewidth}{!}{%
\begin{tabular}{@{}l c cc c c c@{}}
\toprule
Geometry & Method & RelMSE($\boldsymbol{u}$) & PSNR($\boldsymbol{u}$)
       & RelMSE($p$) & Dice($\Omega_t$) & RelMSE($\boldsymbol{d}$) \\
\midrule
\multirow{3}{*}{ConvCH}
& Noisy    & 0.8304 & 32.5829 & /      & 75.6451 & /      \\
& PINN     & 0.1066 & 41.5808 & 0.1722 & 94.0794 & 0.9167 \\
& PirateNet & 0.0243 & 49.5541 & 0.0291 & 98.1288 & 0.2193 \\
\midrule
\multirow{3}{*}{Aorta}
& Noisy    & 0.5643 & 36.8230 & /      & 79.8246 & /      \\
& PINN     & 0.1305 & 44.5081 & 0.2174 & 94.7855 & 1.2618 \\
& PirateNet & 0.0372 & 48.7037 & 0.0558 & 97.5042 & 0.3372 \\
\bottomrule
\end{tabular}}
\end{table}

Table~\ref{tab:main} shows that PirateNet improves velocity,
pressure, displacement, and domain accuracy over the standard PINN in
both geometries. Its random Fourier feature modules may help capture
the finer spatial variations in the flow fields. Velocity RelMSE is only $0.0243$ for ConvCH and $0.0372$
for the aorta, with pressure RelMSE $0.0291$ and $0.0558$,
respectively. These small errors show successful reconstruction in
both the idealized tube and the aortic flow with stronger spatial variation,
without a prescribed initial velocity or inlet profile.
Figures~\ref{fig:convCH_slice} and~\ref{fig:convCH_3d} show noise
suppression and the deforming domain in ConvCH, while
Figures~\ref{fig:aorta_slice} and~\ref{fig:aorta_3d} show the more
varied flow through the curved aortic arch and branches. The aorta
raises velocity RelMSE from $0.1066$ to $0.1305$ for the standard PINN,
compared with $0.0243$ to $0.0372$ for PirateNet. The absolute
increases in velocity, pressure, and displacement errors are larger for
the PINN. On
the aorta, PirateNet's velocity, pressure, and displacement errors remain
$3.5$--$3.9$ times smaller than those of the PINN, which suggests that
PirateNet better represents the stronger spatial variation of the aortic
flow. Pointwise wall displacement is harder to recover than lumen shape:
its small magnitude makes relative errors more pronounced.

\begin{table}
\centering
\caption{Computational cost.}
\label{tab:convergence}
\resizebox{0.6\linewidth}{!}{%
\begin{tabular}{@{}l ccc@{}}
\toprule
Geometry & Outer iterations & Time per iteration & Total time \\
\midrule
ConvCH & 5 & 41.83 \text{mins} & 3.49 \text{hours}\\ 
Aorta & 8 & 41.46 \text{mins} & 5.53 \text{hours}\\ 
\bottomrule
\end{tabular}}
\end{table}

\begin{figure}
    \centering
    \includegraphics[width=\linewidth]{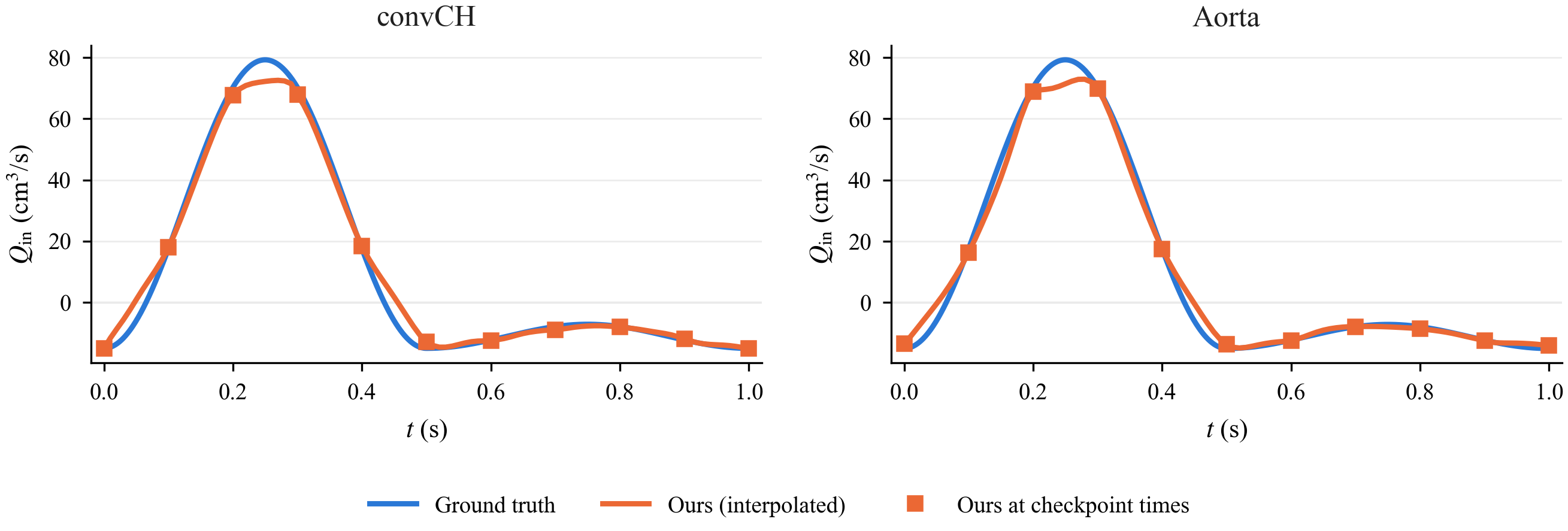}
    \caption{Inlet flux inferred from the learned velocity field
    compared with ground truth for ConvCH and the aorta. Squares mark
    checkpoint times; the orange curves interpolate the reconstructed
    values. No ground-truth inlet profile is supplied during training.}
    \label{fig:inlet}
\end{figure}
Figure~\ref{fig:inlet} presents inlet profile recovery. The inlet
trace is inferred from interior observations and physics residuals. In both geometries, the inferred flux follows the timing
and sign of the ground-truth waveform well, including its rapid rise,
reversal near the main pulse, and weaker later phase, with some
underestimation at the peak.

Table~\ref{tab:convergence} reports outer iterations and wall-clock
costs. Per-iteration times are similar, so the aorta's longer total
runtime mainly reflects its additional outer updates.

\subsection{Effect of the Physics-Based Structure Regularizer}
\label{sec:exp-coupling}

\begin{figure}
    \centering
    \includegraphics[width=\linewidth]{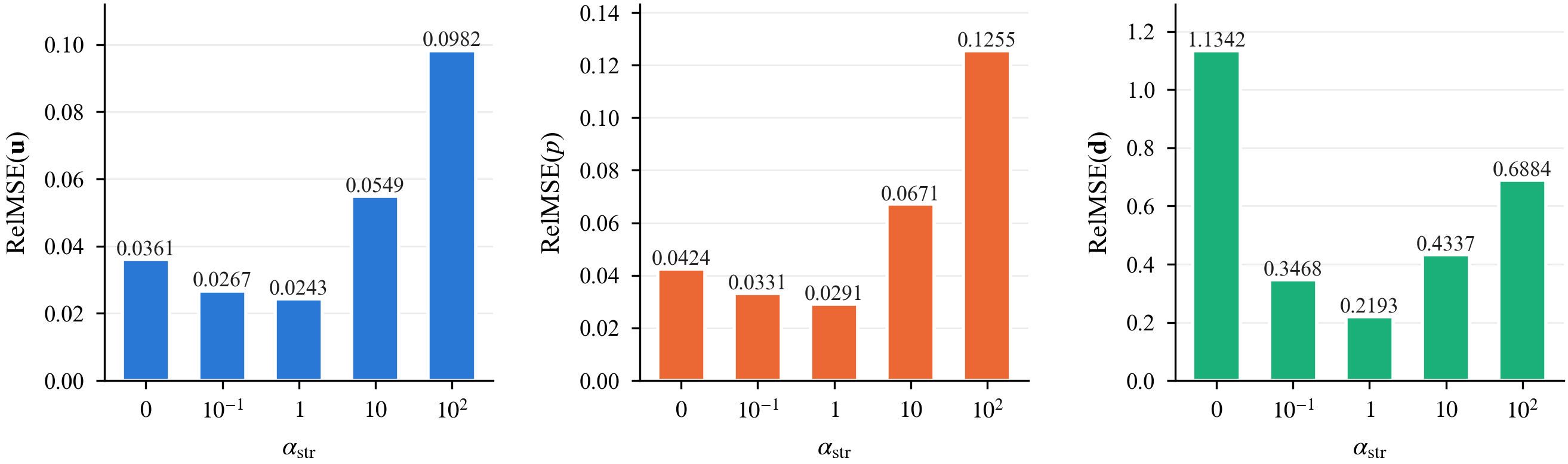}
    \caption{Effect of the structure-regularization weight
    $\alpha_{\mathrm{str}}$ on ConvCH relative errors in velocity,
    pressure, and wall displacement.}
    \label{fig:ablation_sc}
\end{figure}
\begin{table}
\centering
\caption{ConvCH, baseline noise: effect of wall modeling and of the structure regularizer.}
\label{tab:coupling-ablation}
\resizebox{0.9\linewidth}{!}{
\begin{tabular}{@{}l cc c c c@{}}
\toprule
Method & RelMSE($\boldsymbol{u}$) & PSNR($\boldsymbol{u}$)
       & RelMSE($p$) & Dice($\Omega_t$) & RelMSE($\boldsymbol{d}$) \\
\midrule
No-deform $\boldsymbol{d}$      & 0.4164 & 40.4041 & 0.3161 & 94.2875 & /      \\
$\alpha_{\mathrm{str}}=0$       & 0.0361 & 47.2181 & 0.0424 & 98.0834 & 1.1342 \\
$\alpha_{\mathrm{str}}=10^{-1}$ & 0.0267 & 49.2907 & 0.0331 & 98.3714 & 0.3468 \\
$\alpha_{\mathrm{str}}=1$       & 0.0243 & 49.5541 & 0.0291 & 98.1288 & 0.2193 \\
$\alpha_{\mathrm{str}}=10$      & 0.0549 & 45.0228 & 0.0671 & 97.1415 & 0.4337 \\
$\alpha_{\mathrm{str}}=10^{2}$  & 0.0982 & 42.1284 & 0.1255 & 95.2278 & 0.6884 \\
\bottomrule
\end{tabular}}
\end{table}

Table~\ref{tab:coupling-ablation} and Figure~\ref{fig:ablation_sc}
compare fixed-domain and deformable-domain reconstructions at the
baseline ConvCH noise level. The fixed-domain reconstruction has a
velocity RelMSE of $0.4164$, demonstrating the importance of accounting
for wall motion. Allowing the domain to deform without structure
regularization ($\alpha_{\mathrm{str}}=0$) raises Dice to $98.0834\%$,
but displacement RelMSE remains $1.1342$. Accurate lumen overlap
therefore does not, by itself, determine the motion of individual wall
points.

Structure regularization markedly improves pointwise displacement:
its RelMSE falls to $0.3468$ at $\alpha_{\mathrm{str}}=10^{-1}$ and
$0.2193$ at $\alpha_{\mathrm{str}}=1$, while Dice remains near $98\%$.
This improvement suggests that the physical residuals favor wall
correspondences closer to the ground truth, even when the reconstructed
lumen changes little. The $\alpha_{\mathrm{str}}=1$ also yields the lowest velocity
and pressure errors, whereas $\alpha_{\mathrm{str}}=10^{-1}$ gives the
slightly higher Dice of $98.3714\%$. At weights of $10$ and $10^2$,
all three field errors increase and Dice decreases relative to the
$\alpha_{\mathrm{str}}=1$ case, indicating that over-regularization can degrade the overall performance.

\subsection{Sensitivity to Misspecified Physical Parameters}
\label{sec:exp-sensitivity}

We assess sensitivity to imperfect physical coefficients using
the same baseline noisy data and reconstruction setup throughout
Table~\ref{tab:sens} with ConvCH geometry. One sweep perturbs only viscosity $\nu$; the
other changes $\lambda_{\mathrm s}$ and $\mu_{\mathrm s}$ by the same
percentage, preserving their ratio. Both include relative changes of
$0$, $\pm5\%$, $\pm10\%$, and $\pm20\%$.

\begin{figure}
    \centering
    \includegraphics[width=\linewidth]{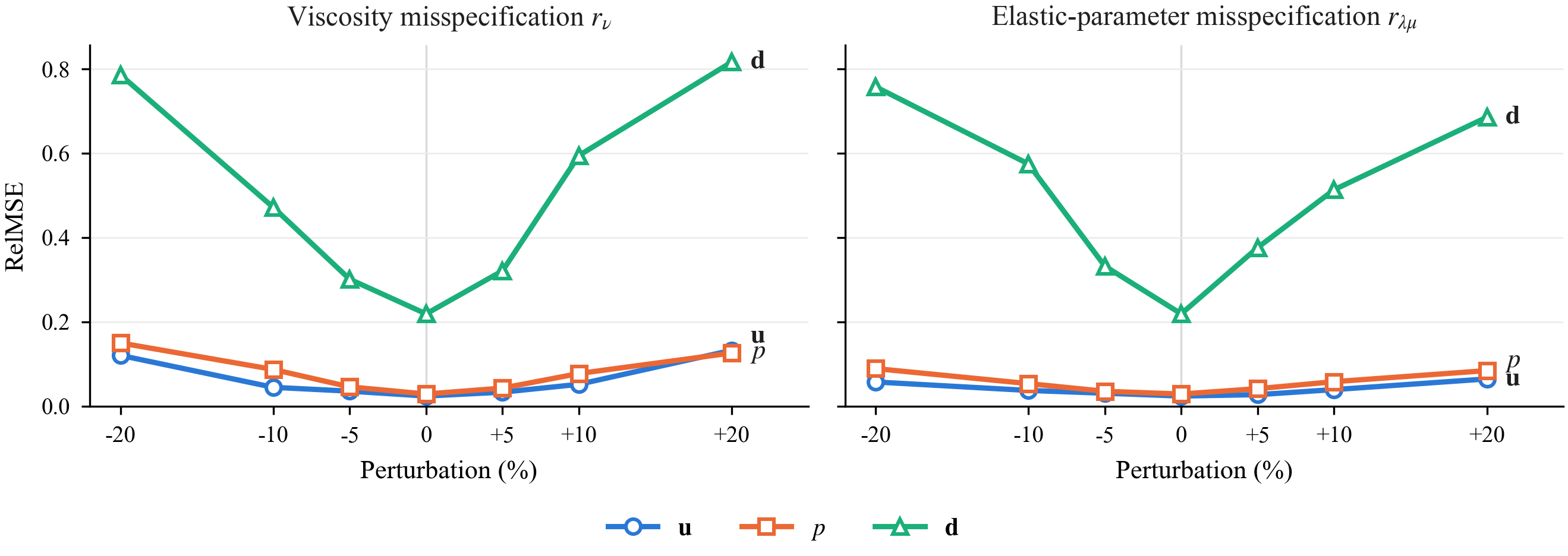}
    \caption{ConvCH reconstruction RelMSE under relative
    perturbations of viscosity (left) or simultaneous proportional
    perturbations of the two Lam\'e coefficients (right).}
    \label{fig:sensitivity}
\end{figure}

\begin{table}
\centering
\caption{Converging-channel reconstruction accuracy under viscosity
and proportional Lam\'e-parameter perturbations at the baseline
noise level. Perturbations are relative to the nominal values.}
\label{tab:sens}
\resizebox{0.9\linewidth}{!}{
\begin{tabular}{@{}l ccccc@{}}
\toprule
\multicolumn{6}{@{}l}{\emph{Viscosity misspecification} $r_\nu$} \\
\midrule
Perturbation & RelMSE($\boldsymbol{u}$) & PSNR($\boldsymbol{u}$)
             & RelMSE($p$) & Dice($\Omega_t$) & RelMSE($\boldsymbol{d}$) \\
\midrule
$-$20\% & 0.1208 & 41.2184 & 0.1503 & 95.1165 & 0.7865 \\
$-$10\% & 0.0450 & 46.0614 & 0.0874 & 97.0328 & 0.4721 \\
$-$5\%  & 0.0359 & 47.3465 & 0.0461 & 97.8752 & 0.3015 \\
0\%     & 0.0243 & 49.5541 & 0.0291 & 98.1288 & 0.2193 \\
$+$5\%  & 0.0335 & 47.6378 & 0.0432 & 97.9415 & 0.3215 \\
$+$10\% & 0.0522 & 45.2253 & 0.0778 & 96.4598 & 0.5952 \\
$+$20\% & 0.1329 & 40.6288 & 0.1265 & 94.5271 & 0.8171 \\
\bottomrule
\toprule
\multicolumn{6}{@{}l}{\emph{Elastic-parameter misspecification} $r_{\lambda\mu}$} \\
\midrule
Perturbation & RelMSE($\boldsymbol{u}$) & PSNR($\boldsymbol{u}$)
             & RelMSE($p$) & Dice($\Omega_t$) & RelMSE($\boldsymbol{d}$) \\
\midrule
$-$20\% & 0.0577 & 44.7189 & 0.0895 & 95.9752 & 0.7582 \\
$-$10\% & 0.0376 & 46.9584 & 0.0538 & 97.6452 & 0.5751 \\
$-$5\%  & 0.0310 & 48.0991 & 0.0356 & 97.9817 & 0.3327 \\
0\%     & 0.0243 & 49.5541 & 0.0291 & 98.1288 & 0.2193 \\
$+$5\%  & 0.0276 & 48.7649 & 0.0418 & 97.8725 & 0.3765 \\
$+$10\% & 0.0396 & 46.7039 & 0.0584 & 97.4575 & 0.5138 \\
$+$20\% & 0.0649 & 44.0902 & 0.0847 & 96.9818 & 0.6861 \\
\bottomrule
\end{tabular}}
\end{table}

Table~\ref{tab:sens} and Figure~\ref{fig:sensitivity} show the lowest
errors at nominal coefficients. Under $\pm5\%$ perturbations,
velocity RelMSE remains between $0.0276$ and $0.0359$ and domain Dice
stays above $97.8\%$ in both sweeps. Even at $\pm10\%$, velocity RelMSE
does not exceed $0.0522$ and Dice remains above $96.4\%$. The method
therefore retains useful flow and domain accuracy over this range of
coefficient misspecification. Pointwise wall displacement is less
stable: its RelMSE rises from $0.2193$ at the nominal setting to
$0.3015$--$0.3765$ under $\pm5\%$ changes. Larger perturbations
produce clearer degradation, especially in viscosity: at $+20\%$,
velocity RelMSE reaches $0.1329$ for viscosity, compared with $0.0649$
for the elastic coefficients. Nevertheless, domain Dice stays above
$94.5\%$ even at $\pm20\%$ in both sweeps. These results show
encouraging tolerance to imperfect fluid and wall coefficients in this
synthetic setting: flow and lumen reconstruction remain accurate under
moderate perturbations, while the wall-displacement estimates benefit
more strongly from coefficients close to their nominal values.

\subsection{Reconstruction with a Windkessel Outlet}
\label{sec:windkessel}
\begin{figure}
    \centering
    \includegraphics[width=\linewidth]{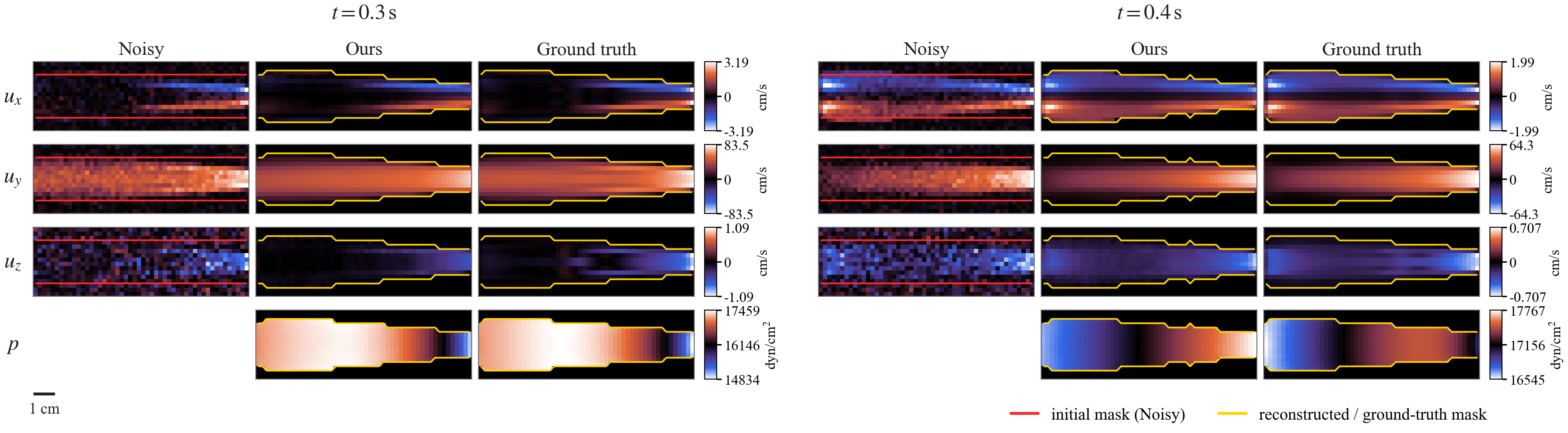}
    \caption{Converging-channel velocity components and pressure at
    $t=0.2$ and $0.3\ \mathrm{s}$ with Windkessel outlet boundary condition. Each time point compares the noisy
    image, reconstruction, and ground truth. Red outlines show the
    initial mask; yellow outlines show the reconstructed and
    ground-truth domains.}
    \label{fig:convCHRCR_slice}
\end{figure}
\begin{figure}[htbp]
    \centering
    \includegraphics[width=\linewidth]{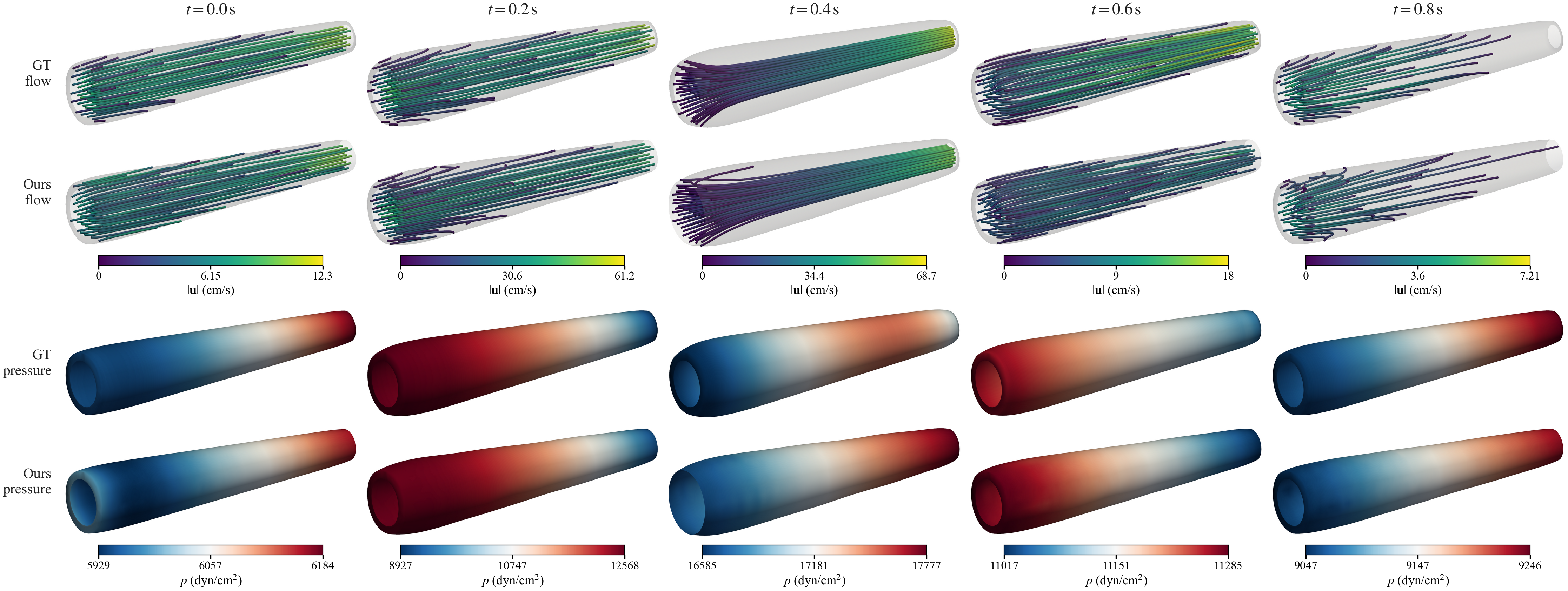}
    \caption{Ground-truth and reconstructed ConvCH flow streamlines
    and wall pressure with a three-element Windkessel outlet condition.
    At each time, the pressure color bar range covers the actual min and max values, since the plotted pressures are all positive.}
    \label{fig:convCHRCR_3d}
\end{figure}

To test a more physiologically realistic outlet setting, we use a
three-element Windkessel (RCR)
condition~\cite{vignon2006outflow,vignonclementel2010outflow} for
ConvCH in place of the baseline traction-free condition
$\boldsymbol\sigma_{\mathrm f}\boldsymbol n_{\mathrm o}=\boldsymbol 0$. We synthesize the noisy data here
using the FEM result computed with this Windkessel condition and the
same noise model as in~\eqref{eq:noise}. The
inverse fluid loss replaces the traction-free outlet term $\mathcal L_{\mathrm{out}}$ with
\begin{equation}
  \mathcal L_{\mathrm{RCR}}
  =\left\|
    \frac{\hat P}{R_{\mathrm d}}
    +C\,\frac{\partial }{\partial t}\hat P
    -\left(1+\frac{R_{\mathrm p}}{R_{\mathrm d}}\right)\hat Q
    -R_{\mathrm p}C\,\frac{\partial }{\partial t}\hat Q
  \right\|_{\mathcal D_{\mathrm o}}^2.
  \label{eq:exp-rcr-loss}
\end{equation}
The RCR loss uses the same outlet weight $\alpha_{\mathrm{out}}$.
Here $\mathcal D_{\mathrm o}$ is the sampled outlet set and
$\hat Q(t)=\int_{\Gamma_t^{\mathrm o}}
\hat{\boldsymbol{u}}\cdot\boldsymbol{n}_{\mathrm o}\,\mathrm dS$
is the outward flow rate. The mean outlet pressure
$\hat P(t)$ is in physical units: the network's
density-normalized pressure $\hat p$ is multiplied by
$\rho_{\mathrm f}$ before outlet averaging. We approximate $\hat Q(t)$ by the outlet area times the sample mean of
$\hat{\boldsymbol u}\cdot\boldsymbol n_{\mathrm o}$, and $\hat P(t)$
by the sample mean of $\rho_{\mathrm f}\hat p$.
We use
$R_{\mathrm p}=100$ and $R_{\mathrm d}=900$ in
$\mathrm{dyn\,s/cm^5}$ and $C=0.001\ \mathrm{cm^5/dyn}$.
Table~\ref{tab:convCH_wks} and Figures~\ref{fig:convCHRCR_slice}--\ref{fig:convCHRCR_3d}
report the resulting reconstruction.

\begin{table}[H]
\centering
\caption{Reconstruction accuracy for the ConvCH with Windkessel outlet boundary condition.}
\label{tab:convCH_wks}
\resizebox{0.9\linewidth}{!}{%
\begin{tabular}{@{}l c cc c c c@{}}
\toprule
Geometry & Method & RelMSE($\boldsymbol{u}$) & PSNR($\boldsymbol{u}$)
       & RelMSE($p$) & Dice($\Omega_t$) & RelMSE($\boldsymbol{d}$) \\
\midrule
\multirow{2}{*}{ConvCH}
& Noisy     & 0.8159 & 38.0788 & /      & 80.8006 & /      \\
& PirateNet & 0.0305 & 52.8383 & 0.0553 & 97.6518 & 0.4046 \\
\bottomrule
\end{tabular}}
\end{table}

The same Gauss--Seidel-type alternating fluid and geometry updates apply in this case, while
the outlet residual changes to match the Windkessel model. The reconstructed velocity has RelMSE $0.0305$
against $0.8159$ for the noisy image, and the moving-domain Dice is
$97.6518\%$. Pressure and wall-displacement RelMSE are $0.0553$ and
$0.4046$, respectively. These results show that the framework can
accommodate a different outlet law while retaining its reference-domain
correction, registration, and structure regularization. Other
differentiable outlet models can be easily incorporated by replacing the
corresponding outlet residual.

\FloatBarrier
\section{Conclusion}
\label{sec:conclusion}

We reconstruct unsteady three-dimensional blood flow, pressure, and
vessel-wall motion from noisy velocity images with strong spatial flow
variation, without prescribed initial velocity or inlet profile. Our
meshless Gauss--Seidel-type alternating framework uses structure regularization to improve
the alignment of pointwise wall displacement with ground truth, as shown
by the ablation study. The reconstruction retains accurate flow and
domain estimates under moderate perturbations of viscosity and elastic
coefficients. The framework readily accommodates different outlet conditions, as
demonstrated by incorporating a more physiologically realistic
three-element Windkessel model through a modified outlet residual.

\section*{Acknowledgements}
The work of Xue-Cheng Tai is partially supported by NORCE Kompetanseoppbygging program. The work of Raymond H. Chan is partially supported by HKRGC Grants CityU11309922, LU13300125, ITF Grant No. MHP/054/22, and LU BGR105824. We gratefully remember our coauthor and dear friend, Professor Alessandro Veneziani, whose insight and substantial contributions were central to this work. We dedicate this paper to his memory.
\bibliographystyle{siamplain}
\bibliography{references}

\clearpage
\appendix
\section{Proofs for the Geometry Results}
\label{app:proofs}

\subsection{Wall Correspondence}
\label{app:proof-wall}

\begin{proof}\,
\smallskip\noindent\textbf{Step 1: Construct a nontrivial wall flow.}\par
Take a chart $\varphi:U\subset\mathbb R^2\to\Gamma_0^{\mathrm w}$
whose image lies strictly inside $S$. Choose
$0\ne\chi\in C_c^\infty(U)$, a scalar bump function, and let
$\boldsymbol e_1=(1,0)^\top$ be the first coordinate vector in $U$.
Since $\operatorname{supp}\chi\Subset U$, its zero extension to
$\mathbb R^2$ is smooth. Set
$\boldsymbol\tau(\varphi(\boldsymbol y))
=D\varphi(\boldsymbol y)(\chi(\boldsymbol y)\boldsymbol e_1)$ on the
chart and $\boldsymbol\tau=\boldsymbol 0$ elsewhere on the wall.
This is a smooth tangential field because it vanishes near the chart
edge. The parameter $s\in\mathbb R$ controls how far points move along
the wall. Consider the ODE
\[
  \frac{\mathrm d}{\mathrm ds}H_s(\boldsymbol y)
    =\chi(H_s(\boldsymbol y))\boldsymbol e_1,
  \qquad H_0(\boldsymbol y)=\boldsymbol y.
\]
It has a smooth flow for all $s$ because its right-hand side has bounded
derivative and is globally Lipschitz. The fundamental theorem on
flows~\cite{lee2012smooth} gives
$H_{s+r}=H_s\circ H_r$ and $H_0=\operatorname{id}$; taking $r=-s$
gives $H_s^{-1}=H_{-s}$. Since the vector field is zero outside
$\operatorname{supp}\chi$, $H_s$ fixes every point there. Since $H_s$
is a bijection of $\mathbb R^2$, it therefore maps
$\operatorname{supp}\chi$ onto itself; thus $H_s(U)=U$. Define
$h_s=\varphi\circ H_s\circ\varphi^{-1}$ on $\varphi(U)$ and set
$h_s$ equal to the identity elsewhere on $\Gamma_0^{\mathrm w}$.
It is a bijection of the entire wall, smooth with smooth inverse
$h_{-s}$ on its smooth part, and equal to the identity near the inlet
and outlet junctions.

\smallskip\noindent\textbf{Step 2: Wall reparameterizations preserve the deformed region.}\par
Choose a smooth function $\psi$ with $\psi(0)=0$ and
$\psi(t_*)\ne0$ at some $t_*>0$. For sufficiently small
$\varepsilon$, put $s=\varepsilon\psi(t)$ and
$\tilde b_t^{\varepsilon}:=b_t\circ h_{\varepsilon\psi(t)}$.
Because $h_{\varepsilon\psi(t)}$ maps the reference wall onto itself,
\[
  \tilde b_t^{\varepsilon}(\Gamma_0^{\mathrm w})
  =b_t\big(h_{\varepsilon\psi(t)}(\Gamma_0^{\mathrm w})\big)
  =b_t(\Gamma_0^{\mathrm w})
  =\Gamma_t^{\mathrm w}.
\]
The construction gives $\tilde b_0^{\varepsilon}=b_0$ and
$\tilde b_t^{\varepsilon}=b_t$ on
$\Gamma_0^{\mathrm w}\setminus S$, hence fixes the inlet and outlet
junctions. The same boundary sets enclose the same lumen region for
every $\varepsilon$.

\smallskip\noindent\textbf{Step 3: A monotone projection distinguishes material-point displacements.}\par
Choose $\boldsymbol x_0=\varphi(\boldsymbol y_0)$ with
$\chi(\boldsymbol y_0)\ne0$. Because $\varphi$ is a chart,
$\boldsymbol v_0:=D\varphi(\boldsymbol y_0)
(\chi(\boldsymbol y_0)\boldsymbol e_1)$ is a nonzero vector in
$T_{\boldsymbol x_0}\Gamma_0^{\mathrm w}$.
Since $b_{t_*}$ is a diffeomorphism,
$b_{t_*}^{-1}\circ b_{t_*}
=\operatorname{id}_{\Gamma_0^{\mathrm w}}$.
Differentiate this identity at $\boldsymbol x_0$ in the tangential
direction $\boldsymbol v_0$. The chain rule gives
\[
  D(b_{t_*}^{-1}\circ b_{t_*})(\boldsymbol x_0)
      \boldsymbol v_0
  =D(b_{t_*}^{-1})(b_{t_*}(\boldsymbol x_0))
      D b_{t_*}(\boldsymbol x_0)\boldsymbol v_0
  =\boldsymbol v_0.
\]
If $D b_{t_*}(\boldsymbol x_0)\boldsymbol v_0$ were zero, the middle
expression would be zero, contradicting $\boldsymbol v_0\ne
\boldsymbol 0$. Hence
$D b_{t_*}(\boldsymbol x_0)\boldsymbol v_0\ne\boldsymbol 0$.
The ODE in Step~1 gives
$\left.\partial_s h_s(\boldsymbol x_0)\right|_{s=0}
=\boldsymbol\tau(\boldsymbol x_0)=\boldsymbol v_0$.
Therefore
$\left.\partial_\varepsilon
 h_{\varepsilon\psi(t_*)}(\boldsymbol x_0)
 \right|_{\varepsilon=0}=\psi(t_*)\boldsymbol v_0$.
Now differentiate
$\tilde b_{t_*}^{\varepsilon}(\boldsymbol x_0)
=b_{t_*}(h_{\varepsilon\psi(t_*)}(\boldsymbol x_0))$ at
$\varepsilon=0$:
\[
  \left.\frac{\mathrm d}{\mathrm d\varepsilon}
     \tilde b_{t_*}^{\varepsilon}(\boldsymbol x_0)
  \right|_{\varepsilon=0}
  =\psi(t_*)\,D b_{t_*}(\boldsymbol x_0)
     \boldsymbol v_0\ne\boldsymbol 0.
\]
Projecting the curve
$\varepsilon\mapsto\tilde b_{t_*}^{\varepsilon}(\boldsymbol x_0)$
onto the direction of its nonzero derivative gives a strictly
monotone scalar function near $0$. Hence distinct small
$\varepsilon$ give distinct displacement fields, while Step~2 keeps
the observed geometry fixed.
\end{proof}

\subsection{Partial Optimality of Alternating Reconstruction}
\label{app:proof-alternating}

\begin{proof}[Proof of Theorem~\ref{thm:alternating-limit}]
\smallskip\noindent\textbf{Step 1: Identify the limiting ALE deformation.}\par
By assumption, $\mathcal G_0^*$ and $\mathcal M_t^*$ are $C^1$
diffeomorphisms of $\overline D$ onto itself. Their composition
$\mathcal A_t^*=\mathcal M_t^*\circ(\mathcal G_0^*)^{-1}$ is therefore
a $C^1$ diffeomorphism. In particular, it maps the limiting reference
domain $\mathcal G_0^*(\Omega_0^0)$ onto the limiting moving
domain $\mathcal M_t^*(\Omega_0^0)$. Continuity of
$\mathcal S_{\mathrm f}$ and convergence of the inlet, initial, and
geometry data give
$(\boldsymbol u^*,p^*)=\mathcal S_{\mathrm f}
[\boldsymbol g^*,\boldsymbol u_0^*;
 \mathcal G_0^*,\boldsymbol d^*]$.

\smallskip\noindent\textbf{Step 2: Show that relative geometry updates vanish.}\par
Since $\mathcal G_0^{n-1}$ maps $\overline D$ onto itself, every
$\boldsymbol y\in\overline D$ can be written as
$\boldsymbol y=\mathcal G_0^{n-1}(\boldsymbol x)$. Therefore
\[
  \|\mathcal C_0^n-\boldsymbol I\|_{L^\infty(\overline D)}
  =\sup_{\boldsymbol x\in\overline D}
    |\mathcal G_0^n(\boldsymbol x)-\mathcal G_0^{n-1}(\boldsymbol x)|
  \longrightarrow0.
\]
Likewise, $\mathcal M_t^{n-1}$ is onto $\overline D$, so
\[
  \sup_{t\in[0,T]}
  \|\mathcal C_t^n-\boldsymbol I\|_{L^\infty(\overline D)}
  =\sup_{t\in[0,T]}\sup_{\boldsymbol x\in\overline D}
    |\mathcal M_t^n(\boldsymbol x)-\mathcal M_t^{n-1}(\boldsymbol x)|
  \longrightarrow0.
\]
These are inter-iteration relative maps. By the definition of the
cumulative map, $\mathcal C_0^n=\mathcal F_0^n$, so the correction made
in each late iteration tends to the identity even when the cumulative
limit $\mathcal G_0^*$ is nontrivial.

\smallskip\noindent\textbf{Step 3: Pass the reference-correction inequality to the limit.}\par
Fix any $\boldsymbol\zeta\in N_0$. The first inequality in
\eqref{eq:asymptotic-stage-optimality}, convergence of
$(\boldsymbol\zeta^n,\boldsymbol u^{n-1},
\mathcal G_0^{n-1})\to(\boldsymbol 0,\boldsymbol u^*,
\mathcal G_0^*)$, continuity of $E_0$,
and $\varepsilon_{0,n}\to0$ give
\[
  E_0(\boldsymbol 0;\boldsymbol u^*,\mathcal G_0^*)
  \le E_0(\boldsymbol\zeta;\boldsymbol u^*,\mathcal G_0^*).
\]
Because the competitor was arbitrary in $N_0$, the limiting
relative correction $\boldsymbol 0$ is a partial local minimizer of
$E_0$ on the limiting initial domain, with the fluid velocity fixed.

\smallskip\noindent\textbf{Step 4: Pass the wall-deformation inequality to the limit.}\par
For an arbitrary $\boldsymbol d\in N_t$, use the second inequality
in~\eqref{eq:asymptotic-stage-optimality}. The fluid state converges
to $(\boldsymbol u^*,p^*)$, while the updated cumulative reference
map, relative correction, preceding displacement, and updated
displacement converge to $\mathcal G_0^*$, $\boldsymbol I$,
$\boldsymbol d^*$, and $\boldsymbol d^*$, respectively.
Continuity of $E_t$ and $\varepsilon_{t,n}\to0$ yield
\[
  E_t(\boldsymbol d^*;\boldsymbol u^*,p^*,\mathcal G_0^*,
      \boldsymbol I,\boldsymbol d^*)
  \le E_t(\boldsymbol d;\boldsymbol u^*,p^*,\mathcal G_0^*,
      \boldsymbol I,\boldsymbol d^*).
\]
This conclusion concerns the full Geometry~2 objective, including
registration, hyperelastic and Laplacian energies,
stress-continuity regularization, and the initial-displacement
penalty.

\smallskip\noindent\textbf{Step 5: Pass the final fluid inequality to the limit.}\par
For any $(\boldsymbol g,\boldsymbol u_0)\in N_{\mathrm f}$,
the third inequality in~\eqref{eq:asymptotic-stage-optimality},
continuity of $E_{\mathrm f}$, and
$\varepsilon_{\mathrm f,n}\to0$ give
\[
  E_{\mathrm f}(\boldsymbol g^*,\boldsymbol u_0^*;
              \mathcal G_0^*,\boldsymbol d^*)
  \le E_{\mathrm f}(\boldsymbol g,\boldsymbol u_0;
                  \mathcal G_0^*,\boldsymbol d^*).
\]
Thus the limiting fluid solution is partially locally optimal among
solutions generated by inlet and initial data in $N_{\mathrm f}$ on
the limiting moving domain. Together with Steps~3--4, this proves
\eqref{eq:alternating-partial-minima}.
\end{proof}

\end{document}